\documentclass[journal]{IEEEtran}
\usepackage{amsmath, amssymb, bm, cite, epsfig, psfrag}
\usepackage{graphicx}
\usepackage{soul}
\usepackage[margin=0.42in]{geometry}
\usepackage{dblfloatfix}
\usepackage{array}
\usepackage{lipsum}
\newcolumntype{P}[1]{>{\centering\hspace{0pt}}p{#1}}
\newcolumntype{M}[1]{>{\centering\hspace{0pt}}m{#1}}
\newcolumntype{L}{>{\centering\arraybackslash}m{3cm}}
\usepackage[font=small]{caption}
\usepackage{epstopdf}	
\usepackage{longtable}
\usepackage{supertabular,booktabs}
\usepackage{bbm}
\usepackage{multirow}
\usepackage[usenames,dvipsnames]{xcolor}
\renewcommand{\arraystretch}{1.5}
\usepackage{subfigure}
\usepackage{etoolbox}
\usepackage{pbox}
\usepackage{fixltx2e}
\usepackage{tabu}	
\usepackage{filecontents}

\usepackage{colortbl}
\usepackage{fancyhdr}

\usepackage{bm}
\newtoggle{conference}
\togglefalse{conference} 
\graphicspath{{figures/}}

\fancypagestyle{firststyle}{
	\fancyhf{}
	\fancyhead[C]{Conference on}     
	\fancyfoot[L]{This is a notice}

}
\begin{document}
\bibliographystyle{IEEEtran}
\title{Overview of Millimeter Wave Communications for Fifth-Generation (5G) Wireless Networks-with a focus on Propagation Models}
\author{\IEEEauthorblockN{Theodore S. Rappaport, \textit{Fellow}, \textit{IEEE}, Yunchou Xing, \textit{Student Member}, \textit{IEEE}, George R. MacCartney, Jr., \textit{Student Member}, \textit{IEEE}, Andreas F. Molisch, \textit{Fellow}, \textit{IEEE}, Evangelos Mellios, \textit{Member}, \textit{IEEE}, Jianhua Zhang, \textit{Senior Member}, \emph{IEEE},\\}\vspace{-0.7cm}
\thanks{T. S. Rappaport (email: tsr@nyu.edu), Y. Xing (email: yx775@nyu.edu), G. R. MacCartney	, Jr. (email: gmac@nyu.edu), are with NYU WIRELESS Research Center, and are supported in part by the NYU WIRELESS Industrial Affiliates: AT\&T, CableLabs, Cablevision, Ericsson, Huawei, Intel Corporation, InterDigital Inc., Keysight Technologies, L3 Communications, Nokia, National Instruments, Qualcomm Technologies, SiBeam, Straight Path Communications, OPPO, Sprint, Verizon and UMC, in part by the GAANN Fellowship Program, and in part by the National Science Foundation under Grant 1320472, Grant 1237821, and Grant 1302336. NYU Tandon School of Engineering,  9th Floor, 2 MetroTech Center,  Brooklyn, NY 11201.}
\thanks{A. F. Molisch (email: molisch@usc.edu), is with the Ming Hsieh Department of Electrical Engineering, University of Southern California, Los Angeles, CA 90089. His work is supported by the National Science Foundation and Samsung.} 
\thanks{E. Mellios (email: Evangelos.Mellios@bristol.ac.uk ), is with the Communication Systems \& Networks Group, University of Bristol, Merchant Venturers Building, Woodland Road, BS8 1UB, Bristol, United Kingdom}
\thanks{J. Zhang (email:jhzhang@bupt.edu.cn), is with State Key Lab of Networking and Switching Technology, Beijing University of Posts and Telecommunications, Mailbox NO.92, 100876.}
\thanks{The authors thank Shu Sun of NYU for her suggestions on this paper.}}

\maketitle
\thispagestyle{fancy}

\begin{abstract}
This paper provides an overview of the features of fifth generation (5G) wireless communication systems now being developed for use in the millimeter wave (mmWave) frequency bands. Early results and key concepts of 5G networks are presented, and the channel modeling efforts of many international groups for both licensed and unlicensed applications are described here. Propagation parameters and channel models for understanding mmWave propagation, such as line-of-sight (LOS) probabilities, large-scale path loss, and building penetration loss, as modeled by various standardization bodies, are compared over the 0.5-100 GHz range.
\end{abstract}
    

\section{Introduction}~\label{sec:intro}

Wireless data traffic has been increasing at a rate of over 50\% per year per subscriber, and this trend is expected to accelerate over the next decade with the continual use of video and the rise of the Internet-of-Things (IoT) \cite{gubbi2013internet,rapfcc16}. To address this demand, the wireless industry is moving to its fifth generation (5G) of cellular technology that will use millimeter wave (mmWave) frequencies to offer unprecedented spectrum and multi-Gigabit-per-second (Gbps) data rates to a mobile device \cite{rappaport2013millimeter}. Mobile devices such as cell phones are typically referred to as user equipment (UE). A simple analysis illustrated that 1 GHz wide channels at 28 or 73 GHz could offer several Gbps of data rate to UE with modest phased array antennas at the mobile handset \cite{rangan2014millimeter}, and early work showed 15 Gbps peak rates are possible with $4 \times 4$ phased arrays antenna at the UE and 200 m spacing between base stations (BSs) \cite{ghosh2014mmwave,roh2014millimeter}. 

Promising studies such as these led the US  Federal Communications Commission (FCC) to authorize its 2016 ``Spectrum Frontiers'' allocation of 10.85 GHz of millimeter wave spectrum for 5G advancements \cite{FCC16-89}, and several studies \cite{singh2015tractable,sundaresan2016fluidnet,banelli2014modulation,michailow2014generalized} have proposed new mobile radio concepts to support 5G mobile networks.

5G mmWave wireless channel bandwidths will be more than ten times greater than today's 4G Long-Term Evolution (LTE) 20 MHz cellular channels. Since the wavelengths shrink by an order of magnitude at mmWave when compared to today's 4G microwave frequencies, diffraction and material penetration will incur greater attenuation, thus elevating the importance of line-of-sight (LOS) propagation, reflection, and scattering. Accurate propagation models are vital for the design of new mmWave signaling protocols (e.g., air interfaces). Over the past few years, measurements and models for a vast array of scenarios have been presented by many companies and research groups \cite{rangan2014millimeter,rappaport2013millimeter,5GCM,haneda2016indoor,deng201528,rappaport201573,
haneda20165g,nie201372,haneda2016frequency,JSAC,rappaport2015wideband,maccartney2015indoor,
maccartney2013path,samimi2015probabilistic,Mac16c,sun2016millimeter,maccartney2015exploiting,
thomas2016prediction,Sun16b,samimi20163,hur2014syn,rappaport2013broadband,Koymen15a}.

This invited overview paper is organized as follows: Section \ref{sec:II} summarizes key 5G system concepts of emerging mmWave wireless communication networks and Section \ref{sec:III} presents 5G propagation challenges and antenna technologies. Section \ref{sec:IV} gives a thorough compilation and comparison of recent mmWave channel models developed by various groups and standard bodies, while Section \ref{sec:conc} provides concluding remarks.

\section{5G System Concepts and Air Interfaces}~\label{sec:II}
\begin{figure*}
    \centering
    \includegraphics[width=0.8\textwidth]{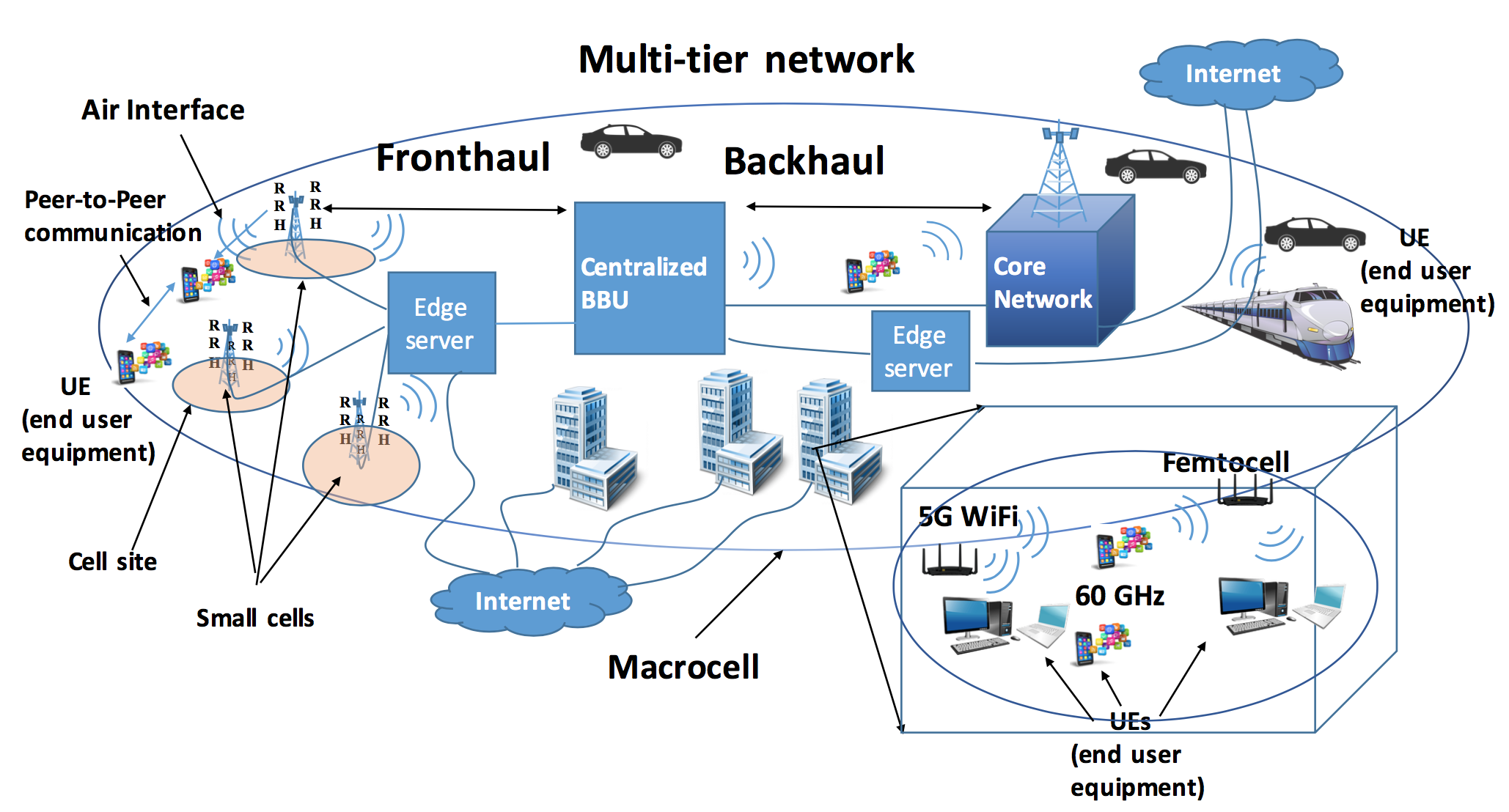}
    \caption{Mobile networks are evolving from 4G towards 5G. Shown here are small cells, edge servers, wireless backhaul, and 5G multi-tier architecture.}
    \label{fig:network}
\end{figure*}

5G promises great flexibility to support a myriad of Internet Protocol (IP) devices, small cell architectures, and dense coverage areas. Applications envisioned for 5G include the Tactile Internet \cite{fettweis2014tactile}, vehicle-to-vehicle communication \cite{mecklenbrauker2011vehicular}, vehicle-to-infrastructure communication \cite{gozalvez2012ieee}, as well as peer-to-peer and machine-to-machine communication \cite{bhushan2014network}, all which will require extremely low network latency and on-call demand for large bursts of data over minuscule time epochs \cite{maeder2011challenge}. Current 4G LTE and WiFi roundtrip latencies are about 20-60 ms \cite{nikravesh2016depth,deng2014wifi}, but 5G will offer roundtrip latencies on the order of 1 ms \cite{andrews2014will}. As shown in Fig. \ref{fig:network}, today's 4G cellular network is evolving to support 5G, where WiFi off-loading, small cells, and distribution of wideband data will rely on servers at the edges of the network (edge servers) to enable new use cases with lower latency.
\subsection{Backhaul and Fronthaul}
Fig. \ref{fig:network} shows how backhaul connects the fixed cellular infrastructure (e.g., BSs) to the core telephone network and the Internet. Backhaul carries traffic between the local subnetwork (e.g., the connections between UE and BSs) and the core network (e.g., the Internet and the Mobile Switching Telephone Office). 4G and WiFi backhaul, and not the air interface, are often sources of traffic bottlenecks in modern networks since backhaul connections provided by packet-based Ethernet-over-Fiber links typically provide only about 1 Gbps \cite{backhaul2015}, which may be easily consumed by several UEs. In a typical macrocell site, a baseband unit (BBU) is in an enclosure at the base of a remote cell site and is directly connected to the backhaul. The BBU processes and modulates IP packet data from the core network into digital baseband signals where they are transmitted to remote radio heads (RRHs). The digital baseband signal travels from the BBU to a RRH via a common public radio interface (CPRI) through a digital radio-over-fiber (D-RoF) connection, also known as fronthaul. The RRH converts the digital signal to analog for transmission over the air at the carrier frequency by connecting to amplifiers and antennas to transmit the downlink from the cell tower. The RRH also converts the received radio frequency (RF) uplink signal from the UEs into a digital baseband signal which travels from the RRH to the BBU via the same CPRI and D-RoF connection to the base of the cell tower. The BBU then processes and packetizes the digital baseband signal from the RRH and sends it through a backhaul connection to the core network. In summary, fronthaul is the connection between the RRH and BBU in both directions and backhaul is the connection between the BBU and the core network in both directions. 

Modern cellular architectures support a more flexible deployment of radio resources that may be distributed using a cloud radio access network technique, where a BS is split into two parts \cite{carapellese2014energy}, one part where the RRHs are at remote cell sites, and in the other part, one centralized BBU is located up to tens of kilometers away (see Fig. \ref{fig:network}). CPRI is used for fronthaul, and interconnects the centralized BBU and multiple RRHs through D-RoF. MmWave wireless backhaul and fronthaul will offer fiber-like data rates and bandwidth to infrastructure without the expense of deploying wired backhaul networks or long-range D-RoF \cite{hur2013millimeter,sundaresan2016fluidnet,H2020}.
\subsection{Small Cells}
\pagestyle{empty}
An effective way to increase area spectral efficiency is to shrink cell size \cite{chandrasekhar2008femtocell,andrews2014will,dohler2011phy} where the reduced number of users per cell, caused by cell shrinking, provides more spectrum to each user. Total network capacity vastly increases by shrinking cells and reusing the spectrum, and future nomadic BSs and direct device-to-device connections between UEs are envisioned to emerge in 5G for even greater capacity per user \cite{wang2014cellular}. Femtocells that can dynamically change their connection to the operator's core network will face challenges such as managing RF interference and keeping timing and synchronization, and various interference avoidance and adaptive power control strategies have been suggested \cite{chandrasekhar2008femtocell}. An analysis of the wireless backhaul traffic at 5.8 GHz, 28 GHz, and 60 GHz in two typical network architectures showed that spectral efficiency and energy efficiency increased as the number of small cells increased \cite{ge20145g}, and backhaul measurements and models at 73 GHz were made in New York City \cite{george2014backhaul,rappaport2015wideband}. Work in \cite{murdock2014consumption} showed a theory for power consumption analysis, which is strikingly similar to noise figure, for comparing energy efficiency and power consumption in wideband networks. An early small-cell paper \cite{haider2011spectral} gave insights into enhancing user throughput, reducing signaling overhead, and reducing dropped call likelihoods.  \vspace{-0.2cm}

\subsection{Multi-tier Architecture}
The roadmap for 5G networks will exploit a multi-tier architecture of larger coverage 4G cells with an underlying network of closer-spaced 5G BSs as shown in Fig. \ref{fig:network}. A multi-tier architecture allows users in different tiers to have different priorities for channel access and different kinds of connections (e.g., macrocells, small cells, and device-to-device connections), thus supporting higher data rates, lower latencies, optimized energy consumption, and interference management by using resource-aware criteria for the BS association and traffic loads allocated over time and space \cite{hossain2014evolution}. Schemes and models for load balanced heterogeneous networks in a multi-tier architecture are given in \cite{andrews2014an,tehrani2014device}. 5G applications will also require novel network architectures that support the convergence of different wireless technologies (e.g., WiFi, LTE, mmWave, low-power IoT) that will interact in a flexible and seamless manner using Software Defined Networking and Network Virtualization principles \cite{yang2015software,agyapong2014design}. 

\subsection{5G Air Interface}
The design of new physical layer air interfaces is an active area of 5G research.  Signaling schemes that provide lower latency, rapid beamforming and synchronization, with much smaller time slots and better spectral efficiency than the orthogonal frequency division multiplexing (OFDM) used in 4G, will emerge. A novel modulation that exploits the dead time in the single-carrier frequency domain modulation method used in today's 4G LTE uplink is given in \cite{ghosh2014mmwave}. Work in \cite{banelli2014modulation} reviews linear modulation schemes such as filter bank multicarrier (FBMC) modulation wherein subcarriers are passed through filters that suppress sidelobes. Generalized frequency division multiplexing (GFDM) is proposed in  \cite{michailow2014generalized}, where it is shown that, when compared with OFDM used in current 4G LTE (which has one cyclic prefix per symbol and high out-of-band emissions \cite{van2008out}), GFDM improves the spectral efficiency and has approximately 15 dB weaker out-of-band emissions. Orthogonal time-frequency-space (OTFS) modulation that spreads the signals in the time-frequency plane has also been suggested, due to superior diversity and higher flexibility in pilot design \cite{monk2016otfs}. Channel state feedback and management to support directional beam search/steering will also be vital \cite{sun2014mimo,haneda2015channel}.

\subsection{5G Unlicensed WiFi}\label{5GWIFI} 
MmWave WiFi for the 57-64 GHz unlicensed bands has been in development for nearly a decade, with the WirelessHD and IEEE 802.11ad standardization process beginning in 2007, and 2009, respectively \cite{Rap15a}. IEEE 802.11ad devices, which can reach 7 Gbps peak rates \cite{yamada2015experimental}, and WirelessHD products which can reach 4 Gbps with theoretical data rates as high as 25 Gbps \cite{siligaris2011a}, are both already available in the market. Building on the history of WiFi standard IEEE 802.11n \cite{charfi2013phy,perahia2013next}, two newer standards, IEEE 802.11ac and 802.11ad, are amendments that improve the throughput to reach 1 Gbps in the 5 GHz band and up to 7 Gbps in the 60 GHz band, respectively. An overview of IEEE Gigabit wireless local area network (WLAN) amendments (IEEE 802.11ac and 802.11ad) \cite{verma2013wifi,perahia2011gigabit,perahia2010ieee} shows the suitability of these two standards for multi-gigabit communications. For the 802.11ad standard \cite{802.11ad}, notable features include fast session transfer for seamless data rate fall back (and rate rise) between 60 GHz and 2.4/5 GHz PHYs, and media access control (MAC) enhancements for directional antennas, beamforming, backhaul, relays and spatial reuse techniques. For enhancements of the PHY layer, beamforming using directional antennas or antenna arrays is used to overcome the increased loss at 60 GHz \cite{Rap15a}. IEEE 802.11ay standard is an ongoing project with the goal to support a maximum throughput of at least 20 Gbps in the 60 GHz unlicensed band \cite{maltsev2016channel}. Newer WiFi standards are sure to emerge to exploit the new 64-71 GHz unlicensed spectrum in the US \cite{FCC16-89}.

\subsection{Vehicular Networks}
Vehicle-to-vehicle (V2V) communications are an important tool for increasing road safety and reducing traffic congestion. Currently the most investigated system is the IEEE 802.11p standard which works in 5.9 GHz band for V2V and vehicle-to-infrastructure (V2I) communication, and is known as dedicated short-range communications (DSRC) \cite{802.11p}. The mmWave bands (e.g., 24  GHz and 77 GHz \cite{FCC16-89}) are attractive for V2V and V2I, (e.g., cars, high-speed railway and subway systems) since connected vehicles will need Gbps date rates, which cannot be achieved in the 10 MHz channel bandwidths at 5.9 GHz in current 4G \cite{moustafa2009vehicular,bendor2011mmwave,rappaport2010analysis}. Limitations of V2V connectivity include the difficulty in achieving realistic spatial consistency to sustain the data-link connection for high-speed mobility vehicles \cite{5GCM,rap2016ap}. Evaluations have shown that narrow beam directional antennas are more suitable for IEEE 802.11p-based systems \cite{shivaldova2012roadside}, and several schemes aimed at utilizing adaptive antennas for fast moving V2V communications are provided in \cite{phan2015making}.

\section{5G Antenna and Propagation Challenges}~\label{sec:III}
The entire radio spectrum up to 5.8 GHz that has been used for global wireless communications throughout the past 100 years easily fits within the bandwidth of the single 60 GHz unlicensed band, yet there is so much more spectrum still available above 60 GHz \cite{FCC16-89,rangan2014millimeter,Rap15a}, as shown in Figure C.1 on page 40 of \cite{Rap15a}. With radio frequency integrated circuits (RFIC) now routinely manufactured for 24 and 77 GHz vehicular radar, and IEEE 802.11ad WiGig devices now becoming mainstream in high-end laptops and cellphones, low-cost electronics will be viable for the evolution of massively broadband 5G millimeter wave communications \cite{rappaport2011state}.

\begin{figure}    
    \centering
    \includegraphics[width=0.40\textwidth]{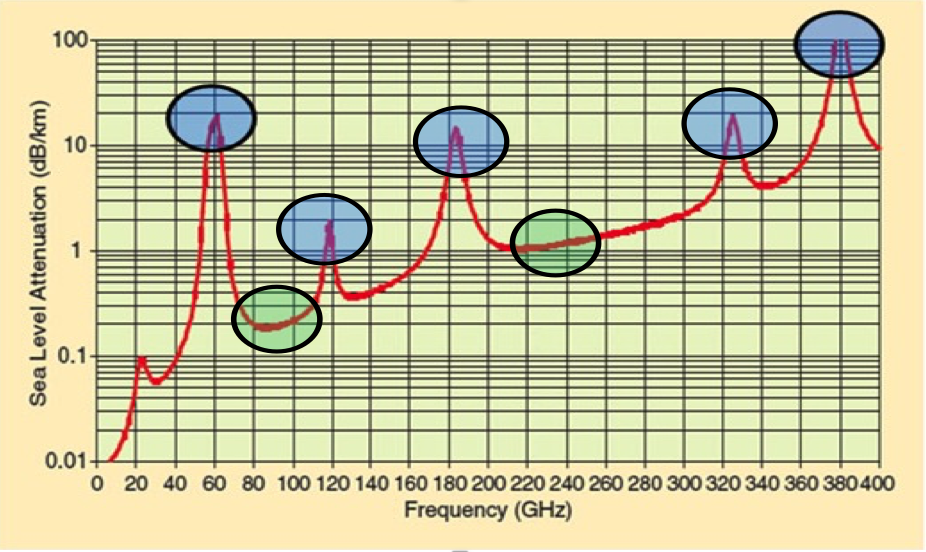}
    \caption{Atmospheric absorption of electromagnetic waves at sea level versus frequency, showing the additional path loss due to atmospheric absorption ~\cite{rappaport2011state}. }
    \label{fig:myths}
\end{figure}

Today, most spectrum above 30 GHz is used for military applications or deep-space astronomy reception, but the recent FCC Spectrum Frontiers ruling has assigned many bands for mobile and backhaul communications. The various resonances of oxygen and other gasses in air, however, cause certain bands to suffer from signal absorption in the atmosphere. Fig. \ref{fig:myths} illustrates how the bands of 183 GHz, 325 GHz, and especially 380 GHz suffer much greater attenuation over distance due to the molecular resonances of various components of the atmosphere, beyond the natural Friis' free space loss, making these particular bands well suited for very close-in communications and ``whisper radio'' applications where massive bandwidth channels will attenuate very rapidly out to a few meters or fractions of a meter \cite{rappaport2013millimeter,Rap15a}. Fig. \ref{fig:myths} also shows many mmWave bands only suffer 1-2 dB more loss than caused by free space propagation per km in air \cite{ITU-Rattenuation,sun2017a}. Rain and hail cause substantial attenuation at frequencies above 10 GHz \cite{xu2000measurements}, and 73 GHz signals attenuate at 10 dB/km for a 50 mm/hr rain rate  \cite{rappaport2013millimeter,ITU-Rspecific,Rap15a}. Interestingly, as shown in \cite{rappaport2013millimeter,rappaport2011state} rain attenuation flattens out at 100 GHz to 500 GHz, and for all mmWave frequencies, rain or snow attenuation may be overcome with additional antenna gain or transmit power. Also, the size and orientation of rain drops and clouds may determine the particular amount of attenuation on air-to-ground links such that satellites could undergo more localized and perhaps less rain attenuation than terrestrial links at mmWave frequencies.  

While it is commonly believed that path loss increases dramatically by moving up to mmWave frequencies, extensive work in various environments in \cite{Sun16b,maccartney2015indoor,sun2016propagation,samimi20163,sun2015path}  shows that Friis' equation \cite{friis1946note} dictates this is true only when the antenna gain is assumed to be constant over frequency. If the physical size of the antenna (e.g., effective aperture) is kept constant over frequency at both link ends and the weather is clear, then path loss in free space actually \textit{decreases} quadratically as frequency \textit{increases} \cite{Rap15a}. The larger antenna gains at higher frequencies require adaptive beam steering for general use at both the BS and UE, compared to legacy mobile antennas with lower gain \cite{Rap15a}. Beam steerable antenna technologies estimate directions of arrival and adaptively switch beam patterns to mitigate interference and to capture the signal of interest. Adaptive arrays are essential for mmWave communications to compensate the path loss caused by  blockage from dynamic obstacles \cite{Rap15a,uchendu2016survey,nitsche2015steering,samimi20163,sun2017a,maccartney2016millimeter}.

Penetration into buildings may pose a significant challenge for mmWave communication, and this is a distinct difference from today's UHF/microwave systems. Measurements at 38 GHz described in \cite{rodriguez2015analysis} found a penetration loss of nearly 25 dB for a tinted glass window and 37 dB for a glass door. Measurements at 28 GHz \cite{rappaport2013millimeter} showed that outdoor tinted glass and brick pillars had penetration losses of 40.1 dB and 28.3 dB, respectively, but indoor clear glass and drywall only had 3.6 dB and 6.8 dB of loss. Work in \cite{jacque2016indoor} shows penetration losses for many common materials and provides normalized attenuation (e.g., in dB/cm) at 73 GHz.  

\begin{figure}  
    \centering
    \includegraphics[width=0.40\textwidth]{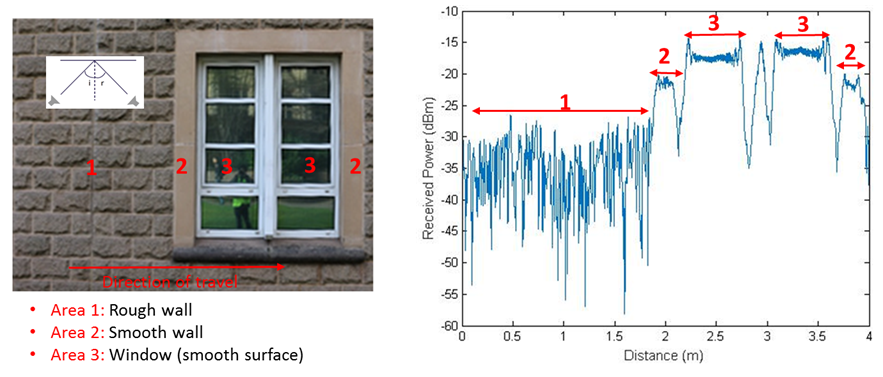}
    \caption{Results of diffuse scattering measurements at 60 GHz, where smooth surfaces (e.g., windows) offer high correlation over distance, but signals from rough surfaces seem less correlated over distance ~\cite{rumney2016testing2,mmMAGIC}. }
    \label{fig:Bristol}
\end{figure}

MmWave will need to exploit and rapidly adapt to the spatial dynamics of the wireless channel since greater gain antennas will be used to overcome path loss. Diffuse scattering from rough surfaces may introduce large signal variations over very short travel distances (just a few centimeters) as shown in Fig. \ref{fig:Bristol}. Such rapid variations of the channel must be anticipated for proper design of channel state feedback algorithms, link adaptation schemes and beam-forming/tracking algorithms, as well as ensuring efficient design of MAC and Network layer transmission control protocols (TCP) that induce re-transmissions. Measurement of diffuse scattering at 60 GHz on several rough and smooth wall surfaces \cite{rumney2016testing2,mmMAGIC} demonstrated large signal level variations in the first order specular and in the non-specular scattered components (with fade depths of up to 20 dB) as a user moved by a few centimeters. In addition, the existence of multipath from nearly co-incident signals can create severe small-scale variations in the channel frequency response. As reported in \cite{rumney2016testing2,mmMAGIC}, measurements showed that reflection from rough materials might suffer from high depolarization, a phenomenon that highlights the need for further investigation into the potential benefits of exploiting polarization diversity for the performance enhancement of mmWave communication systems. Work in \cite{samimi201628} showed shallow Ricean fading of multipath components and exponential decaying trends for spatial autocorrelation at 28 GHz and quick decorrelation at about 2.5 wavelengths for the LOS environment. Work in \cite{rap2016ap} shows that received power of wideband 73 GHz signals has a stationary mean over slight movements but average power can change by 25 dB as the mobile transitioned a building cornor from non-line-of-sight (NLOS) to LOS in an urban microcell (UMi) environment \cite{samimi2016local,maccartney2016millimeter}. Measurements at 10, 20 and 26 GHz demonstrate that diffraction loss can be predicted using well-known models as a mobile moves around a corner using directional antennas \cite{sijia2016}, and human body blockage causes more than 40 dB of fading \cite{maccartney2016millimeter,samimi2016local}.


It is not obvious that the stationarity region size or small-scale statistics derived from 3GPP TR 36.873 \cite{3GPP2014} and other sub-6 GHz channel models, or those used by 3GPP or ITU above 6 GHz are valid for mmWave channels \cite{sun2017a,ertel1998overview,sun2017b,rappaport20175g,rappaport2017VTC}. Recent measurements \cite{rap2016ap,rumney2016testing2,samimi2016local} indicate very sharp spatial decorrelation over small distance movements of just a few tens of wavelengths at mmWave, depending on antenna orientation, but more work is needed in this area. The necessity and proper form of spatial consistency, if borne out by measurements, have yet to be fully understood by the research community.
\section{Channel Modeling}~\label{sec:IV}
\begin{table*}[!ht]\footnotesize
\renewcommand{\arraystretch}{1.1}
\centering
\caption{LOS probability models in the UMi scenario.} \label{tbl:UMiLOS}
\newcommand{\tabincell}[2]{\begin{tabular}{@{}#1@{}}#2\end{tabular}}
\begin{tabular}{|c|c|c|}
\hline
\tabincell{c}{ }&\tabincell{c}{\textbf{LOS probability models (distances are in meters)}}&\tabincell{c}{\textbf{Parameters}}\\
\hline
\tabincell{c}{3GPP TR 38.901\cite{3GPP2017} }&\tabincell{l}{\textbf{Outdoor users:}\\$P_{LOS}(d_{2D}) = \min(d_1/d_{2D},1)(1-\exp(-d_{2D}/d_2))+ \exp(-d_{2D}/d_2)$\\ \textbf{Indoor users:}\\ Use $d_{2D-out}$ in the formula above instead of $d_{2D}$}&\tabincell{c}{$d_1 =18 \;\text{m}$ \\ $d_2= 36 \;\text{m}$}\\
\hline
\tabincell{c}{5GCM\cite{5GCM}}&\tabincell{l}{\textbf{$\bm{d_1/ d_2}$ model:}\\$P_{LOS}(d_{2D}) = \min(d_1/d_{2D},1)(1-\exp(-d_{2D}/d_2))+ \exp(-d_{2D}/d_2)$\\ \textbf{NYU (squared) model:}\\$P_{LOS}(d_{2D}) = (\min(d_1/d_{2D},1)(1-\exp(-d_{2D}/d_2))+ \exp(-d_{2D}/d_2))^2 $}&\tabincell{c}{\textbf{$\bm{d_1/ d_2}$ model:}\\$d_1 =20 \;\text{m}$ \\ $d_2= 39 \;\text{m}$\\ \textbf{NYU (squared) model:}\\$d_1 =22 \;\text{m}$ \\ $d_2= 100 \;\text{m}$ }\\
\hline
\tabincell{c}{METIS\cite{METIS2015} }&\tabincell{l}{\textbf{Outdoor users:}\\$P_{LOS}(d_{2D}) = \min(d_1/d_{2D},1)(1-\exp(-d_{2D}/d_2))+ \exp(-d_{2D}/d_2)$\\ \textbf{Indoor users:}\\ Use $d_{2D-out}$ in the formula above instead of $d_{2D}$}&\tabincell{c}{$d_1 =18 \;\text{m}$ \\ $d_2= 36 \;\text{m}$ \\ $10 \;\text{m} \leq d_{2D}$}\\
\hline
\tabincell{c}{mmMAGIC\cite{mmMAGIC} }&\tabincell{l}{\textbf{Outdoor users:}\\$P(d_{2D}) = \min(d_1/d_{2D},1)(1-\exp(-d_{2D}/d_2))+ \exp(-d_{2D}/d_2)$\\ \textbf{Indoor users:}\\ Use $d_{2D-out}$ in the formula above instead of $d_{2D}$}&\tabincell{c}{$d_1 =20 \;\text{m}$ \\ $d_2= 39 \;\text{m}$}\\
\hline
\multicolumn{3}{|l|}{\textbf{Note :} $d_{2D}$ is the 2D T-R Euclidean distance, and $d_{2D-out}$ is the 2D Euclidean distance of the straight line between the TX and building facade}\\
\hline
\end{tabular}
\end{table*}
\begin{table*}[!ht]\footnotesize
\renewcommand{\arraystretch}{1.1}
\centering
\caption{LOS probability models for the UMa scenario} \label{tbl:UMaLOS}	
\newcommand{\tabincell}[2]{\begin{tabular}{@{}#1@{}}#2\end{tabular}}
\begin{tabular}{|c|c|c|}
\hline
\tabincell{c}{ }&\tabincell{c}{\textbf{LOS probability models (distances are in meters)}}&\tabincell{c}{\textbf{Parameters}}\\
\hline
\tabincell{c}{3GPP TR 38.901\cite{3GPP2017} }&\tabincell{l}{\textbf{Outdoor users:}\\$P_{LOS} = \left( \min(d_1/d_{2D},1)(1-\exp(-d_{2D}/d_2))+ \exp(-d_{2D}/d_2)\right) (1+C(d_{2D},h_{UE}))$\\ where,\\$ C(d_{2D},h_{UE})=\begin{cases}
0, &  h_{UE}      <      13\;\text{m}\\
\left( \frac{h_{UE}-13}{10}\right) ^{1.5}g(d_{2D}),    &   13\;\text{m}\leq h_{UE} \leq 23 \;\text{m}
\end{cases}$ \\ and,\\$ g(d_{2D})=\begin{cases}
0,               &   d_{2D}      \leq      18\;\text{m}\\
(1.25e-6)(d_{2D})^3 \exp(-d_{2D}/150),   &   18\;\text{m} < d_{2D}   
\end{cases}  $\\ \textbf{Indoor users:}\\ Use $d_{2D-out}$ in the formula above instead of $d_{2D}$}&\tabincell{c}{$d_1 =18 \;\text{m}$ \\ $d_2= 63 \;\text{m}$}\\
\hline
\tabincell{c}{5GCM\cite{5GCM}}&\tabincell{l}{\textbf{$\bm{d_1/ d_2}$ model:}\\$P_{LOS} = \left( \min(d_1/d_{2D},1)(1-\exp(-d_{2D}/d_2))+ \exp(-d_{2D}/d_2)\right) (1+C(d_{2D},h_{UE}))$\\ \textbf{NYU (squared) model:}\\$P_{LOS} = \left( \left( \min(d_1/d_{2D},1)(1-\exp(-d_{2D}/d_2))+ \exp(-d_{2D}/d_2)\right) (1+C(d_{2D},h_{UE}))\right) ^2$}&\tabincell{c}{\textbf{\bm{$d_1/ d_2$} model:}\\$d_1 =20 \;\text{m}$ \\ $d_2= 66 \;\text{m}$\\ \textbf{NYU (squared) model:}\\$d_1 =20 \;\text{m}$ \\ $d_2= 160 \;\text{m}$ }\\
\hline
\tabincell{c}{METIS\cite{METIS2015}}&\tabincell{l}{\textbf{Outdoor users:}\\$P_{LOS} = \left( \min(d_1/d_{2D},1)(1-\exp(-d_{2D}/d_2))+ \exp(-d_{2D}/d_2)\right) (1+C(d_{2D},h_{UE}))$\\ \textbf{Indoor users:}\\ Use $d_{2D-out}$ in the formula above instead of $d_{2D}$}&\tabincell{c}{$d_1 =18 \;\text{m}$ \\ $d_2= 63 \;\text{m}$}\\
\hline
\end{tabular}
\end{table*} 

Channel models are required for simulating propagation in a reproducible and cost-effective way, and are used to accurately design and compare radio air interfaces and system deployment. Common wireless channel model parameters include carrier frequency, bandwidth, 2-D or 3-D distance between transmitter (TX) and receiver (RX), environmental effects, and other requirements needed to build globally standardized equipment and systems. The definitive challenge for a 5G channel model is to provide a fundamental physical basis, while being flexible, and accurate, especially across a wide frequency range such as 0.5 GHz to 100 GHz. Recently, a great deal of research aimed at understanding the propagation mechanisms and channel behavior at the frequencies above 6 GHz has been published 
\cite{rangan2014millimeter,rappaport2013millimeter,5GCM,haneda2016indoor,deng201528,rappaport201573,
haneda20165g,nie201372,haneda2016frequency,JSAC,rappaport2015wideband,maccartney2015indoor,
maccartney2013path,samimi2015probabilistic,Mac16c,maccartney2015exploiting,
thomas2016prediction,Sun16b,samimi20163,hur2014syn,rappaport2013broadband,Koymen15a,andrews2014will,
rappaport2011state,sun2016millimeter,sun2016propagation,sun2015path,rodriguez2015analysis,
rumney2016testing2,mmMAGIC,samimi201628,rap2016ap,samimi2016local,sijia2016,3GPP2017,samimi20153,
xu2000measurements,bendor2011mmwave,MiWEBA,hur2016proposal,ITU-RM.2135,jacque2016indoor,METIS2015,jarvelainen2016evaluation,piersanti2012millimeter,semaan2014outdoor,haneda2015channel,maccartney2015millimeter,maccartney2017study}. The specific types of antennas used and numbers of measurements collected vary widely and may generally be found in the referenced work.

For the remainder of this paper, the models for LOS probability, path loss, and building penetration introduced by four major organizations in the past years are reviewed and compared: (i) the 3rd Generation Partnership Project (3GPP TR 38.901 \cite{3GPP2017}), which attempts to provide channel models from 0.5-100 GHz based on a modification of 3GPP's extensive effort to develop models from 6 to 100 GHz in TR 38.900 \cite{3GPP2016}. 3GPP TR documents are a continual work in progress and serve as the international industry standard for 5G cellular, (ii) 5G Channel Model (5GCM) \cite{5GCM}, an ad-hoc group of 15 companies and universities that developed  models based on extensive measurement campaigns and helped seed 3GPP understanding for TR 38.900 \cite{3GPP2016}, (iii) Mobile and wireless communications Enablers for the Twenty-twenty Information Society (METIS) \cite{METIS2015} a large research project sponsored by European Union, and (iv) Millimeter-Wave Based Mobile Radio Access Network for Fifth Generation Integrated Communications (mmMAGIC) \cite{mmMAGIC}, another large research project sponsored by the European Union. While many of the participants overlap in these standards bodies, the final models between those groups are somewhat distinct. It is important to note that recent work has found discrepancies between standardized models and measured results \cite{samimi20163,rappaport2017VTC,rappaport20175g}.


\subsection{LOS Probability Model}
The mobile industry has found benefit in describing path loss for both LOS and NLOS conditions separately. As a consequence, models for the probability of LOS are required, i.e., statistical models are needed to predict the likelihood that a UE is within a clear LOS of the BS, or in an NLOS region due to obstructions. LOS propagation will offer more reliable performance in mmWave communications as compared to NLOS conditions, given the greater diffraction loss at higher frequencies compared to sub-6 GHz bands where diffraction is a dominant propagation mechanism ~\cite{sijia2016,rap2016ap}, and given the larger path loss exponent as well as increased shadowing variance in NLOS as compared to LOS~\cite{Sun16b}. The LOS probability is modeled as a function of the 2D TX-RX (T-R) separation distance and is frequency-independent, as it is solely based on the geometry and layout of an environment or scenario \cite{samimi2015probabilistic}. In the approach of 5GCM \cite{5GCM}, the LOS state is determined by a map-based approach in which only the TX and the RX positions are considered for determining if the direct path between the TX and RX is blocked. 

\subsubsection{UMi LOS Probability}
The UMi scenarios include high user density open areas and street canyons with BS heights below rooftops (e.g., 3-20 m), UE heights at ground level (e.g., 1.5 m) and inter-site distances (ISDs) of 200 m or less \cite{3GPP2014,ITU-RM.2135}. The UMi LOS probability models developed by the various parties are provided in Table \ref{tbl:UMiLOS} and are detailed below.

\paragraph{3GPP TR 38.901} 
The antenna height is assumed to be 10 m in the UMi LOS probability model~\cite{3GPP2017} and the model is referred to as the 3GPP/ITU $d_1/ d_2$ model (it originates in ~\cite{ITU-RM.2135,3GPP2014}), with $d_1$ and $d_2$  curve-fit parameters shown in Table \ref{tbl:UMiLOS}. In \cite{3GPP2017}, model parameters were found to be $d_1=18 \;\text{m}$ and $d_2 = 36 \;\text{m}$ for UMi. For a link between an outdoor BS and an indoor UE, the model uses the outdoor distance $d_{2D-out}$, which is the distance from the BS to the surface of the indoor building, to replace $d_{2D}$. 
        
\paragraph{5GCM}
 5GCM provides two LOS probability models, the first one is identical in form to the 3GPP TR 38.901 outdoor model \cite{3GPP2017}, but with slightly different curve-fit parameters ($d_1$ and $d_2$). The second LOS probability model is the \emph{NYU squared} model \cite{samimi2015probabilistic}, which improves the accuracy of the $d_1/ d_2$ model by including a square on the last term. The NYU model was developed using a much finer resolution intersection test than used by 3GPP TR 38.901, and used a real-world database in downtown New York City \cite{samimi2015probabilistic}. For UMi, the 5GCM $d_1/d_2$ model has a slightly smaller mean square error (MSE), but the \textit{NYU squared} model has a more realistic and rapid decay over distance for urban clutter \cite{5GCM,samimi2015probabilistic}.

\paragraph{METIS}
 The LOS probability model used in METIS \cite{METIS2015} is based on the work of 3GPP TR 36.873 \cite{3GPP2014}, and has the same form and the same parameter values as the 3GPP TR 38.901 model in Table~\ref{tbl:UMiLOS} where the minimum T-R separation distance is assumed to be 10 m in the UMi scenario. 
\paragraph{mmMAGIC}
For the UMi scenario, the mmMAGIC LOS probability model and parameter values are identical to the 5GCM $d_1/d_2$ model \cite{5GCM}. 	
 
\subsubsection{UMa LOS Probability}

Urban macrocell (UMa) scenarios typically have BSs mounted above rooftop levels of surrounding buildings (e.g., 25-30 m) with UE heights at ground level (e.g., 1.5 m) and ISDs no more than 500 m \cite{3GPP2014,ITU-RM.2135}. The UMa LOS probability models are given in Table \ref{tbl:UMaLOS} and are identical to the UMi LOS probability models but with different $d_1$ and $d_2$ values. 

\paragraph{3GPP TR 38.901}
The 3GPP TR 38.901 UMa LOS probability models for outdoor and indoor users are presented in Table~\ref{tbl:UMaLOS}, where for indoor users, $d_{2D-out}$ is used instead of $d_{2D}$ and the models are derived assuming the TX antenna height is 25 m. Due to the larger antenna heights in the UMa scenario, mobile height is an added parameter of the LOS probability as shown in Table~\ref{tbl:UMaLOS} where $h_{UE}$ represents the UE antenna height above ground.

\begin{figure}
    \centering
    \includegraphics[width=0.40\textwidth]{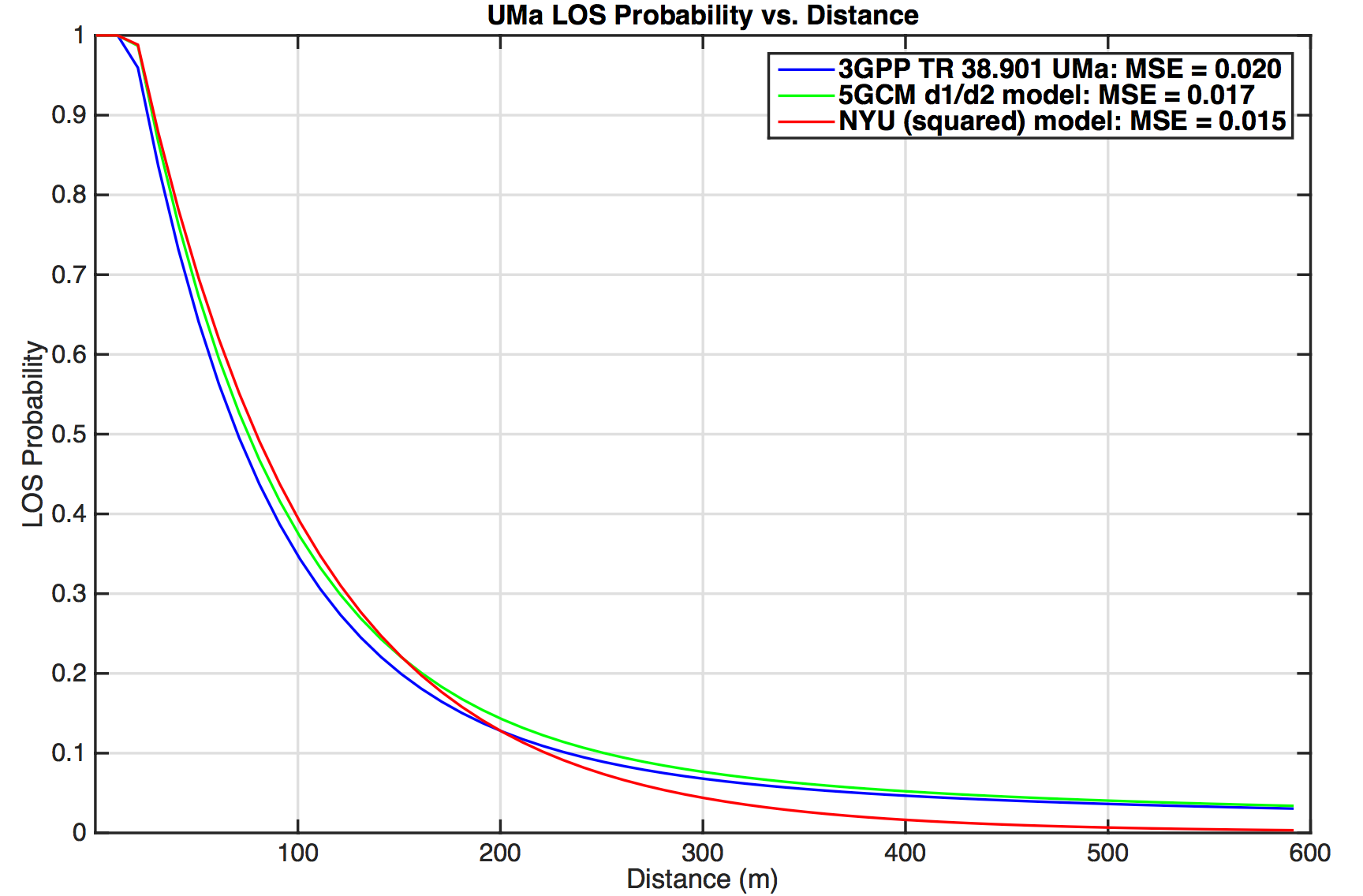}
    \caption{Comparison among three different LOS probability models in UMa scenario.}
    \label{fig:LOSpro}
\end{figure}

\paragraph{5GCM}
The UMa LOS probability models in the 5GCM white paper \cite{5GCM} are of the same form as those in 3GPP TR 38.901 \cite{3GPP2017}, but with different  $d_1$ and $d_2$ values. The 5GCM includes the \emph{NYU squared} option \cite{samimi2015probabilistic}, similar to the UMi scenario. Differences between the 3GPP TR 38.901 and 5GCM UMa LOS probability models are given via MSE in Fig. \ref{fig:LOSpro} for a UE height of 1.5 m. Similar performances are found among the three models, with the \emph{NYU squared} model having the lowest MSE, while also providing the most conservative (e.g., lowest probability) for LOS at distance of several hundred meters \cite{5GCM,samimi2015probabilistic}. 

\paragraph{METIS}
The LOS probability model used in \cite{METIS2015} has the same form as the one in 3GPP TR 38.901 in Table~\ref{tbl:UMaLOS}, and the minimum T-R separation distance is assumed to be 35 m in the UMa scenario.
\paragraph{mmMAGIC}
The UMa scenario is taken into account in the channel model, however, it is not explicitly mentioned in the table since frequency spectrum above 6 GHz is expected to be used for small cell BSs.

\subsubsection{InH LOS Probability}

\begin{table}[!ht]\footnotesize
\renewcommand{\arraystretch}{1.1}
\centering
\caption{LOS probability models in the InH scenario} \label{tbl:InHLOS}
\newcommand{\tabincell}[2]{\begin{tabular}{@{}#1@{}}#2\end{tabular}}
\begin{tabular}{|c|c|}
\hline

\tabincell{c}{\textbf{3GPP TR 38.901\cite{3GPP2017} (all distances are in meters) }}\\
\hline
\tabincell{l}{\textbf{InH-Mixed office:}\\$P_{LOS}=\begin{cases}
1,                & d_{2D} \leq1.2\;\text{m}\\
\exp{(-(d_{2D}-1.2)/4.7)} ,      & 1.2\;\text{m} <d_{2D}      <  6.5\;\text{m}\\
\exp{(-(d_{2D}-6.5)/32.6)}\cdot 0.32,     &  6.5\;\text{m} \leq d_{2D} 
\end{cases}  $\\ \textbf{InH-Open office:}\\ $P_{LOS}^{\text{Open-office}}=\begin{cases}
1,   & d_{2D} \leq 5\;\text{m}\\
\exp{(-(d_{2D}-5)/70.8)} ,   & 5\;\text{m} <d_{2D}      <  49\;\text{m}\\
\exp{(-(d_{2D}-49)/211.7)}\cdot 0.54,    &  49 \;\text{m} \leq d_{2D} 
\end{cases}  $}\\
\hline
\tabincell{c}{\textbf{5GCM\cite{5GCM}}}\\
\hline
\tabincell{l}{$P_{LOS}=\begin{cases}
1,                & d_{2D} \leq1.2\;\text{m}\\
\exp{(-(d_{2D}-1.2)/4.7)} ,      & 1.2\;\text{m} <d_{2D}      <  6.5\;\text{m}\\
\exp{(-(d_{2D}-6.5)/32.6)}\cdot 0.32,     &  6.5\;\text{m} \leq d_{2D} 
\end{cases}  $}\\
\hline
\tabincell{c}{\textbf{mmMAGIC\cite{mmMAGIC}} }\\
\hline
\tabincell{c}{$P_{LOS}=\begin{cases}
1,                & d_{2D} \leq1.2\;\text{m}\\
\exp{(-(d_{2D}-1.2)/4.7)} ,      & 1.2 <d_{2D}      <  6.5\;\text{m}\\
\exp{(-(d_{2D}-6.5)/32.6)}\cdot 0.32,     &  6.5\;\text{m} \leq d_{2D} 
\end{cases}  $}\\
\hline
\end{tabular}
\end{table}

\paragraph{3GPP TR 38.901}
The indoor office environment consists of two types: indoor hotspot (InH)-Mixed office and InH-Open office, where the density of obstructions is greater in the mixed office. LOS probability models for a TX antenna height of 3 m for the InH-Mixed office and InH-Open office sub-scenarios are provided in Table~\ref{tbl:InHLOS}.

\paragraph{5GCM}
In ~\cite{5GCM}, different types of indoor office environments were investigated, including open-plan offices with cubicle areas, closed-plan offices with corridors and meeting rooms, and hybrid-plan offices with both open and closed areas, and based on ray-tracing simulations \cite{jarvelainen2016evaluation}. See Table \ref{tbl:InHLOS} and \cite{5GCM}. 

\paragraph{mmMAGIC}
mmMAGIC adopted the 5GCM InH scenario LOS probability model \cite{5GCM}. 

\subsubsection{RMa LOS Probability}
Rural macrocell (RMa) scenarios typically have BS heights that range between 10 m and 150 m with UE heights at ground level (e.g., 1.5 m) and ISDs up to 5000 m \cite{3GPP2014,ITU-RM.2135}. The LOS probabilities for RMa were not specified in METIS or 5GCM channel models. The 3GPP TR 38.901 \cite{3GPP2017} RMa LOS probability model was adopted from the International Telecommunications Union-Radio (ITU-R) M.2135~\cite{ITU-RM.2135}, which was derived from the WINNER \cite{WINNERplus} RMa LOS probability model and is given by:
\begin{equation}\label{equ:3GPPRmaLOSprob}
\footnotesize
P_{LOS} = \begin{cases}
1, & d_{2D}\leq 10 \text{ m}\\
\exp\left(-\frac{d_{2D}-10}{1000}\right), & d_{2D} > 10\text{ m}
\end{cases}
\end{equation}
where $P_{LOS}$ is the LOS probability for a specific T-R pair, $d_{2D}$ is the 2D T-R separation distance (in meters). Similarly, the RMa LOS probability 3GPP TR 38.901 Release 14 channel model~\cite{3GPP2017} is adopted entirely from ITU-R M.2135~\cite{ITU-RM.2135}. As shown in \cite{JSAC,Mac16c}, caution is advised since these models were derived from urban (not rural) scenarios below 6 GHz.

\subsection{Large-Scale Path Loss Models}

\begin{table*}[!ht]\footnotesize
\renewcommand{\arraystretch}{1.1}
\centering
\caption{Omnidirectional Path loss models in the UMi scenario} \label{tbl:UMiPL}
\newcommand{\tabincell}[2]{\begin{tabular}{@{}#1@{}}#2\end{tabular}}
\begin{tabular}{|c|c|c|c|}
\hline
\tabincell{c}{ }&\tabincell{c}{\textbf{PL [dB], $f_c$ is in GHz and $d_{3D}$ is in meters}}&\tabincell{c}{\textbf{Shadow fading} \\\textbf{std [dB]} }&\tabincell{c}{\textbf{Applicability range} \\ \textbf{and Parameters}}\\
\hline
\multicolumn{4}{|c|}{\textbf{5GCM \cite{5GCM}} }\\
\hline
\tabincell{c}{5GCM UMi-Street \\Canyon LOS}&\tabincell{c}{\textbf{CI model with 1 m reference distance:}\\ $PL = 32.4 + 21 \log_{10}(d_{3D}) + 20 \log_{10}(f_c)$}&\tabincell{c}{$\sigma_{SF} = 3.76$}&\tabincell{c}{$6 < f_c < 100 \;\text{GHz}$}\\
\hline
\tabincell{c}{5GCM UMi-Street \\Canyon NLOS}&\tabincell{c}{\textbf{CI model with 1 m reference distance:}\\ $PL = 32.4 + 31.7 \log_{10}(d_{3D}) + 20 \log_{10}(f_c)$\\ \textbf{ABG model:}\\ $PL = 35.3\log_{10}(d_{3D}) + 22.4 + 21.3\log_{10}(f_c)$}&\tabincell{c}{ \\$\sigma_{SF} = 8.09$\\ \\ $\sigma_{SF} = 7.82$}&\tabincell{c}{$6 < f_c < 100 \;\text{GHz}$}\\
\hline
\tabincell{c}{5GCM UMi-Open \\Square LOS}&\tabincell{c}{\textbf{CI model with 1 m reference distance:}\\ $PL = 32.4 + 18.5\log_{10}(d_{3D}) + 20 \log_{10}(f_c)$}&\tabincell{c}{$\sigma_{SF} = 4.2$}&\tabincell{c}{$6 < f_c < 100 \;\text{GHz}$}\\
\hline
\tabincell{c}{5GCM UMi-Open \\Square NLOS}&\tabincell{c}{\textbf{CI model with 1 m reference distance:}\\ $PL = 32.4 + 28.9 \log_{10}(d_{3D}) + 20 \log_{10}(f_c)$\\ \textbf{ABG model:}\\ $PL = 41.4\log_{10}(d_{3D}) + 3.66 + 24.3\log_{10}(f_c)$}&\tabincell{c}{ \\$\sigma_{SF} = 7.1$\\ \\ $\sigma_{SF} = 7.0$}&\tabincell{c}{$6 < f_c < 100 \;\text{GHz}$}\\
\hline
\multicolumn{4}{|c|}{\textbf{3GPP TR 38.901 V14.0.0 \cite{3GPP2017}} }\\
\hline
\tabincell{c}{3GPP UMi-Street \\Canyon LOS}&\tabincell{c}{$PL_{UMi-LOS}=\begin{cases}
PL_1, & 10\;\text{m} \leq d_{2D} \leq d_{BP}'\\
PL_2,& d_{BP}' \leq d_{2D} \leq 5 \;\text{km}
\end{cases}$\\ $PL_1 = 32.4 + 21\log_{10}(d_{3D}) + 20 \log_{10}(f_c)$\\ $PL_2 = 32.4 + 40 \log_{10}(d_{3D}) + 20\log_{10}(f_c)$ \\
$-9.5\log_{10}((d_{BP}')^2 + (h_{BS}-h_{UE})^2)$\\ where $d_{BP}'$ is specified in Eq. (\ref{equ:UMi3GPPbp}) \ }&\tabincell{c}{$\sigma_{SF} = 4.0$}&\tabincell{c}{$0.5 < f_c < 100 \;\text{GHz}$\\ $1.5 \;\text{m} \leq h_{UE} \leq 22.5 \;\text{m}$\\ $h_{BS} = 10 \;\text{m}$}\\
\hline
\tabincell{c}{3GPP UMi-Street \\Canyon NLOS }&\tabincell{c}{$PL = \max{ \left( PL_{UMi-LOS}(d_{3D}), PL_{UMi-NLOS}(d_{3D}) \right) } $\\ $PL_{UMi-NLOS} = 35.3\log_{10}(d_{3D}) + 22.4 + 21.3\log_{10}(f_c)$\\$ -0.3(h_{UE}-1.5)$\\ \textbf{Option: CI model with 1 m reference distance}\\$PL = 32.4 + 20\log_{10}(f_c)+31.9\log_{10}(d_{3D})$}&\tabincell{c}{\\ $\sigma_{SF} = 7.82$\\ \\$\sigma_{SF} = 8.2$ }&\tabincell{c}{$0.5 < f_c < 100 \;\text{GHz}$\\$ 10 \;\text{m} < d_{2D} < 5000 \;\text{m}$\\ $1.5\;\text{m} \leq h_{UE} \leq 22.5 \;\text{m}$\\ $h_{BS} = 10 \;\text{m}$ }\\
\hline
\multicolumn{4}{|c|}{\textbf{METIS \cite{METIS2015}} }\\
\hline
\tabincell{c}{METIS UMi-Street\\ Canyon LOS }&\tabincell{c}{$PL_{UMi-LOS}\begin{cases}
PL_1, & 10\;\text{m} < d_{3D} \leq d_{BP}\\
PL_2,& d_{BP} < d_{3D} \leq 500 \;\text{m}
\end{cases}$\\ $PL_1 = 22 \log_{10}(d_{3D}) + 28.0
 + 20\log_{10}(f_c) + PL_{0}$\\ $PL_2 = 40 \log_{10}(d_{3D})+7.8 -18 \log_{10}(h_{BS} h_{UE})$\\
 $ + 2\log_{10}(f_c) + PL_1(d_{BP})$\\ $d_{BP}$ and $PL_0$ are specified in  Eq. (\ref{equ:METiSBP}) and (\ref{equ:METiSPL})}&\tabincell{c}{$\sigma_{SF} = 3.1 $ }&\tabincell{c}{$0.8 \leq f_{c} \leq 60 \;\text{GHz}$}\\ 
\hline
\tabincell{c}{METIS UMi-Street\\ Canyon NLOS}&\tabincell{c}{$PL = \max{ \left( PL_{UMi-LOS}(d_{3D}), PL_{UMi-NLOS}(d_{3D}) \right) }$\\$PL_{UMi-NLOS}=36.7\log_{10}(d_{3D}) +23.15 +26\log_{10}(f_c) -0.3(h_{UE})$}&\tabincell{c}{$\sigma_{SF} = 4.0 $ }&\tabincell{c}{$0.45 \leq f_{c} \leq 6 \;\text{GHz}$\\ $ 10 \;\text{m} < d_{2D} <2000 \;\text{m}$ \\ $h_{BS}=10 \;\text{m}$\\$1.5\;\text{m} \leq h_{UE} \leq 22.5 \;\text{m}$}\\
\hline
\multicolumn{4}{|c|}{\textbf{mmMAGIC \cite{mmMAGIC}} }\\
\hline
\tabincell{c}{mmMAGIC UMi-Street \\Canyon LOS}&\tabincell{c}{$PL =  19.2 \log_{10}(d_{3D}) + 32.9 + 20.8\log_{10}(f_c)$}&\tabincell{c}{$\sigma_{SF} = 2.0$}&\tabincell{c}{$6 < f_c < 100 \;\text{GHz}$}\\
\hline
\tabincell{c}{mmMAGIC UMi-Street \\Canyon NLOS}&\tabincell{c}{$PL = 45.0 \log_{10}(d_{3D}) + 31.0 + 20.0 \log_{10}(f_c)$}&\tabincell{c}{ \\$\sigma_{SF} = 7.82$}&\tabincell{c}{$6 < f_c < 100 \;\text{GHz}$}\\
\hline
\multicolumn{4}{|l|}{\textbf{Note :} $PL$ is path loss. $d_{3D}$ is the 3D T-R Euclidean distance.}\\
 \multicolumn{4}{|l|}{\ \ \ \ \ \ \ \ All distances or heights are in meters and frequency related values are in GHz, unless it is stated otherwise.}\\
\hline
\end{tabular}
\end{table*}

There are three basic types of large-scale path loss models to predict mmWave signal strength over distance for the vast mmWave frequency range (with antenna gains included in the link budget and not in the slope of path loss as shown in Eq. (3.9) of \cite{Rap15a}, also see p.3040 in \cite{rappaport2015wideband}). These include the close-in (CI) free space reference distance model (with a 1 m reference distance) \cite{sun2016propagation,rappaport2015wideband,Sun16b,sun2015path}, the  CI model with a frequency-weighted or height weighted path loss exponent (CIF and CIH models) \cite{maccartney2015indoor,haneda2016frequency,Mac16c,JSAC}, and the floating intercept (FI) path loss model, also known as the ABG model because of its use of three parameters $\alpha$, $\beta$, and $\gamma$ \cite{hata1980,haneda2016frequency, maccartney2015indoor, piersanti2012millimeter, maccartney2013path,rappaport2015wideband}. Standard bodies historically create omnidirectional path loss models with the assumption of unity gain antennas for generality. However, it is worth noting that omnidirectional path loss models will not be usable in directional antenna system analysis unless the antenna patterns and true spatial and temporal multipath channel statistics are known or properly modeled \cite{JSAC,rappaport2015wideband,samimi20163,sun2015synthesizing,sun2017a,rappaport20175g,maccartney2014primrc}.
 
The CI path loss model accounts for the frequency dependency of path loss by using a close-in reference distance based on Friis' law as given by \cite{5GCM,Sun16b,maccartney2015indoor,JSAC,Mac16c}:
 \begin{equation}
 \footnotesize
 \label{equ:CI}
\begin{split}
 PL^{CI}(f_c,d_{3D})\;\text{[dB]} = \text{FSPL} (f_c,1 \;\text{m}) + 10n\log_{10}\left( d_{3D} \right)+ \chi_{\sigma}^{CI}
 \end{split}
\end{equation}
where $\chi_{\sigma}^{CI}$ is the shadow fading (SF) that is modeled as a zero-mean Gaussian random variable with a standard deviation in dB, $n$ is the path loss exponent (PLE) found by minimizing the error of the measured data to \eqref{equ:CI}, $d_{3D} > 1m$, $\text{FSPL} (f,1 \;\text{m})$ is the free space path loss (FSPL) at frequency $f_c$ in GHz at 1 m and is calculated by \cite{friis1946note,JSAC}:
\begin{equation}\label{FSPL}
\footnotesize
\begin{split}
\text{FSPL}(f_c , 1\;\text{m})= 20 \log_{10}\left( \frac{4 \pi f_c \times 10^9}{c}\right)=32.4 + 20\log_{10}(f_c)\ \text{[dB]}
\end{split} 
\end{equation}
where $c$ is the speed of light, $3 \times 10^8$ m/s. Using \eqref{FSPL} it is clear that \eqref{equ:CI} can be represented as given in Table \ref{tbl:UMiPL}. The standard deviation $\sigma$ yields insight into the statistical variation about the distant-dependent mean path loss \cite{Rap15a}.

The CI model ties path loss at any frequency to the physical free space path loss at 1 m according to Friis' free space equation \cite{friis1946note}, and has been shown to be robust and accurate in various scenarios \cite{Sun16b,JSAC,Mac16c,thomas2016prediction}. Indoor environments, however, were found to have frequency-dependent loss beyond the first meter, due to the surrounding environment, and work in \cite{maccartney2015indoor} extended the CI model to the CIF model where the PLE has a frequency-dependent term. Recent work \cite{JSAC,Mac16c} has made 73 GHz rural measurements to beyond 10 km and adapted the CIF model form to predict path loss as a function of TX antenna height in RMa scenarios, as path loss was found to be accurately predicted with a height dependency in the PLE, leading to the CIH model\footnote{The CIH model has the same form as \eqref{equ:CIF} except the PLE is a function of the BS height in the RMa scenario instead of frequency, as given by: $PL^{CIH} (f_c,d,h_{BS})~\text{[dB]} = 32.4+20\log_{10}(f_c)+10n\left(1+b_{tx}\left(\frac{h_{BS}-h_{B0}}{h_{B0}}\right)\right)\log_{10}(d)+\chi_{\sigma}, \text{where}~d\geq \text{1 m, and }h_{B0}$ is a reference RMa BS height \cite{JSAC}.}, which has the same form of the CIF model given in \eqref{equ:CIF}:


 \begin{equation}
\footnotesize
\begin{split}
 \label{equ:CIF}
 PL^{CIF}(f_c,d)\;\text{[dB]} &= 32.4 + 20\log_{10}(f_c) \\
&+ 10n\left(1+b\left(\frac{f_c-f_0}{f_0} \right)  \right) \log_{10}\left( d \right) + \chi_{\sigma}^{CIF}
 \end{split}
\end{equation}
where $n$ denotes the distance dependence of path loss, $b$ is an optimization parameter that describes the linear dependence of path loss about the weighted average of frequencies $f_0$ (in GHz), from the data used to optimize the model \cite{maccartney2015indoor,Mac16c,JSAC}.

The CIF model uses two parameters to model average path loss over distance, and reverts to the single parameter CI model when $ b = 0$ for multiple frequencies, or when a single frequency $f=f_0$ is modeled \cite{5GCM,haneda2016indoor,maccartney2015indoor,haneda20165g,JSAC}.

The FI/ABG path loss model is given as:
 \begin{equation}
\footnotesize
 \label{equ:ABG}
 \begin{split}
 PL^{ABG}(f_c,d)\;\text{[dB]} = 10\alpha \log_{10}(d) + \beta + 10\gamma \log_{10}(f_c) + \chi^{ABG}_\sigma
 \end{split}
\end{equation}
where three model parameters $\alpha$, $\beta$ and $\gamma$ are determined by finding the best fit values to minimize the error between the model and the measured data. In (\ref{equ:ABG}), $\alpha$ indicates the slope of path loss with log distance, $\beta$ is the floating offset value in dB, and $\gamma$ models the frequency dependence of path loss, where $f_c$ is in GHz. 

Generalizations of the CI, CIF, and FI/ABG models consider different slopes of path loss over distance before and after a breakpoint distance, where the location of the breakpoint depends mostly on the environment. The dual-slope CIF model is:
\begin{equation}\label{equ:CIFdual}
\scriptsize{
PL_{Dual}^{CIF}(d)\;\text{[dB]}=\begin{cases}
FSPL(f_c , 1\;\text{m}) \\
+ 10n_1\left( 1+b_1\left( \frac{f_c-f_0}{f_0}\right)  \right)\log_{10}( d)  , & 1<d \leq d_{BP}\\
FSPL(f_c , 1\;\text{m}) \\
+ 10n_1\left( 1+b_1\left( \frac{f_c-f_0}{f_0}\right)  \right) \log_{10}(d_{BP})\\
+ 10n_2\left( 1+ b_2\left( \frac{f_c-f_0}{f_0} \right) \right) \log_{10}(\frac{d}{d_{BP}}), & d>d_{BP}
\end{cases}}
\end{equation}
The dual-slope ABG model is:
\begin{equation}\label{equ:ABGdual}
\footnotesize{
PL_{Dual}^{ABG}(d)\;\text{[dB]} =\begin{cases}
\alpha_1\ast 10\log_{10} (d) + \beta_1 \\ 
+ \gamma \ast 10\log_{10}(f_c) , & 1<d \leq d_{BP}\\
\alpha_1\ast 10\log_{10} (d_{BP}) + \beta_1\\
 + \gamma \ast 10\log_{10}(f_c)\\
+\alpha_2 \ast 10 \log_{10}(\frac{d}{d_{BP}}), & d>d_{BP}
\end{cases}}
\end{equation}
where the $\alpha_1$ and $\alpha_2$ are the ``dual slope'' and $d_{BP}$ is the breakpoint distance. Both dual-slope models require 5 parameters to predict distant-dependent average path loss (frequencies are in GHz and distances are in meters).

\begin{table*}[!ht]\footnotesize
\renewcommand{\arraystretch}{1.1}
\centering
\caption{Omnidirectional Path loss models in the UMa scenario} \label{tbl:UMaPL}
\newcommand{\tabincell}[2]{\begin{tabular}{@{}#1@{}}#2\end{tabular}}
\begin{tabular}{|c|c|c|c|}
\hline
\tabincell{c}{ }&\tabincell{c}{\textbf{PL [dB], $f_c$ is in GHz, $d$ is in meters}}&\tabincell{c}{\textbf{Shadow fading} \\\textbf{std [dB]} }&\tabincell{c}{\textbf{Applicability range} \\ \textbf{and Parameters}}\\
\hline
\multicolumn{4}{|c|}{\textbf{5GCM \cite{5GCM}} }\\
\hline
\tabincell{c}{5GCM UMa \\ LOS}&\tabincell{c}{\textbf{ CI model with 1 m reference distance:}\\ $PL = 32.4 + 20 \log_{10}(d_{3D}) + 20 \log_{10}(f_c)$  }&\tabincell{c}{$\sigma_{SF} = 4.1$}&\tabincell{c}{$6 < f_c < 100 \;\text{GHz}$}\\
\hline
\tabincell{c}{5GCM UMa \\ NLOS}&\tabincell{c}{\textbf{CI model with 1 m reference distance:}\\ $PL = 32.4 + 30 \log_{10}(d_{3D}) + 20 \log_{10}(f_c)$\\ \textbf{ABG model:}\\ $PL = 34\log_{10}(d_{3D}) + 19.2 + 23\log_{10}(f_c)$}&\tabincell{c}{ \\$\sigma_{SF} = 6.8$\\ \\ $\sigma_{SF} = 6.5$}&\tabincell{c}{$6 < f_c < 100 \;\text{GHz}$}\\
\hline
\multicolumn{4}{|c|}{\textbf{3GPP TR 38.901 V14.0.0 \cite{3GPP2017}} }\\
\hline
\tabincell{c}{3GPP TR 38.901 UMa \\ LOS}&\tabincell{c}{$PL_{UMa-LOS}=\begin{cases}
PL_1, & 10\;\text{m} \leq d_{2D} \leq d_{BP}'\\
PL_2,& d_{BP}' \leq d_{2D} \leq 5 \;\text{km}
\end{cases}$\\$PL_1 = 28.0 + 22 \log_{10}(d_{3D}) + 20 \log_{10}(f_c)$\\ $PL_2 = 28.0 + 40 \log_{10}(d_{3D}) + 20\log_{10}(f_c)$ \\
$-9\log_{10}((d_{BP}')^2 + (h_{BS}-h_{UE})^2)$\\ where $d_{BP}' = 4 h_{BS}'h_{UE}'f_c \times 10^9/c$ }&\tabincell{c}{$\sigma_{SF} = 4.0$}&\tabincell{c}{$0.5 < f_c < 100 \;\text{GHz}$\\ $1.5 \;\text{m} \leq h_{UE} \leq 22.5 \;\text{m}$\\ $h_{BS} = 25 \;\text{m}$}\\
\hline
\tabincell{c}{3GPP TR 38.901 UMa \\ NLOS }&\tabincell{c}{$PL = \max{ \left( PL_{UMa-LOS}(d_{3D}), PL_{UMa-NLOS}(d_{3D}) \right) } $\\ $PL_{UMa-NLOS} = 13.54+ 39.08\log_{10}(d_{3D}) + 20 \log_{10}(f_c)$\\$ -0.6(h_{UE}-1.5)$\\ \textbf{Option: CI model with 1 m reference distance}\\$PL = 32.4 + 20\log_{10}(f_c)+30 \log_{10}(d_{3D})$}&\tabincell{c}{\\ $\sigma_{SF} = 6.0$\\ \\$\sigma_{SF} = 7.8$ }&\tabincell{c}{$0.5 < f_c < 100 \;\text{GHz}$\\$ 10 \;\text{m} < d_{2D} < 5000 \;\text{m}$\\ $1.5\;\text{m} \leq h_{UE} \leq 22.5 \;\text{m}$\\ $h_{BS} = 25 \;\text{m}$ }\\
\hline
\multicolumn{4}{|c|}{\textbf{METIS \cite{METIS2015}} }\\
\hline
\tabincell{c}{METIS UMa \\ LOS}&\tabincell{c}{$PL_{UMa-LOS}=\begin{cases}
PL_1, & 10\;\text{m} \leq d_{2D} \leq d_{BP}'\\
PL_2,& d_{BP}' \leq d_{2D} \leq 5 \;\text{km}
\end{cases}$\\$PL_1 = 28 + 22 \log_{10}(d_{3D}) + 20 \log_{10}(f_c)$\\ $PL_2 = 28 + 40 \log_{10}(d_{3D}) + 20\log_{10}(f_c)$ \\
$-9\log_{10}((d_{BP}')^2 + (h_{BS}-h_{UE})^2)$\\ where $d_{BP}' = 4 (h_{BS}-1)(h_{UE}-1)f_c \times 10^9 /c$ }&\tabincell{c}{$\sigma_{SF} = 4.0$}&\tabincell{c}{$0.45 < f_c < 6 \;\text{GHz}$\\ $ 10 \;\text{m} < d_{2D} < 5000 \;\text{m}$\\$1.5 \;\text{m} \leq h_{UE} \leq 22.5 \;\text{m}$\\ $h_{BS} = 25 \;\text{m}$}\\
\hline
\tabincell{c}{METIS UMa \\ NLOS }&\tabincell{c}{$PL = \max{ \left( PL_{UMa-LOS}(d_{3D}), PL_{UMa-NLOS}(d_{3D}) \right) } $\\ $PL_{UMa-NLOS} = 161.94 -7.1 \log_{10}(w) + 7.5 \log_{10}(h)$\\ $-\left( 24.37-3.7\left( \dfrac{h}{h_{BS}}\right)^2  \right)\log_{10}(h_{BS})$\\$+(43.42-3.1\log_{10}(h_{BS}))(\log_{10}(d_{3D})-3)$\\$ + 20\log_{10}(f_c)-0.6(h_{UE})$}&\tabincell{c}{\\ $\sigma_{SF} = 6.0$}&\tabincell{c}{$0.45 < f_c < 6 \;\text{GHz}$\\$ 10 \;\text{m} < d_{2D} < 5000 \;\text{m}$\\ $1.5\;\text{m} \leq h_{UE} \leq 22.5 \;\text{m}$\\ $h_{BS} = 25 \;\text{m}$\\$ w = 20 \;\text{m}$\\$h = 20 \;\text{m}$ }\\
\hline
\end{tabular}
\end{table*}

\subsubsection{UMi Large-Scale Path Loss}
\paragraph{5GCM}        
In the 5GCM white paper \cite{5GCM}, the CI model (\ref{equ:CI}) is chosen for modeling UMi LOS path loss, since $\alpha$ in the ABG model (\ref{equ:ABG}) is almost identical to the PLE of the CI model, and also $\gamma$ is very close to 2 which is predicted by the physically-based Friis' free space equation and used in the CI model \cite{Sun16b}. Both the CI and ABG models were adopted for UMi NLOS in 5GCM, and the parameters values for the CI and ABG models are given in Table \ref{tbl:UMiPL}. In the CI path loss model, only a single parameter, the PLE, needs to be determined through optimization to minimize the model error of mean loss over distance, however, in the ABG model,  three parameters need to be optimized to minimize the error, but with very little reduction of the shadowing variance compared to the CI model \cite{sun2016propagation,Sun16b,maccartney2015indoor}. 

 \begin{figure}
    \centering
    \includegraphics[width=0.4\textwidth]{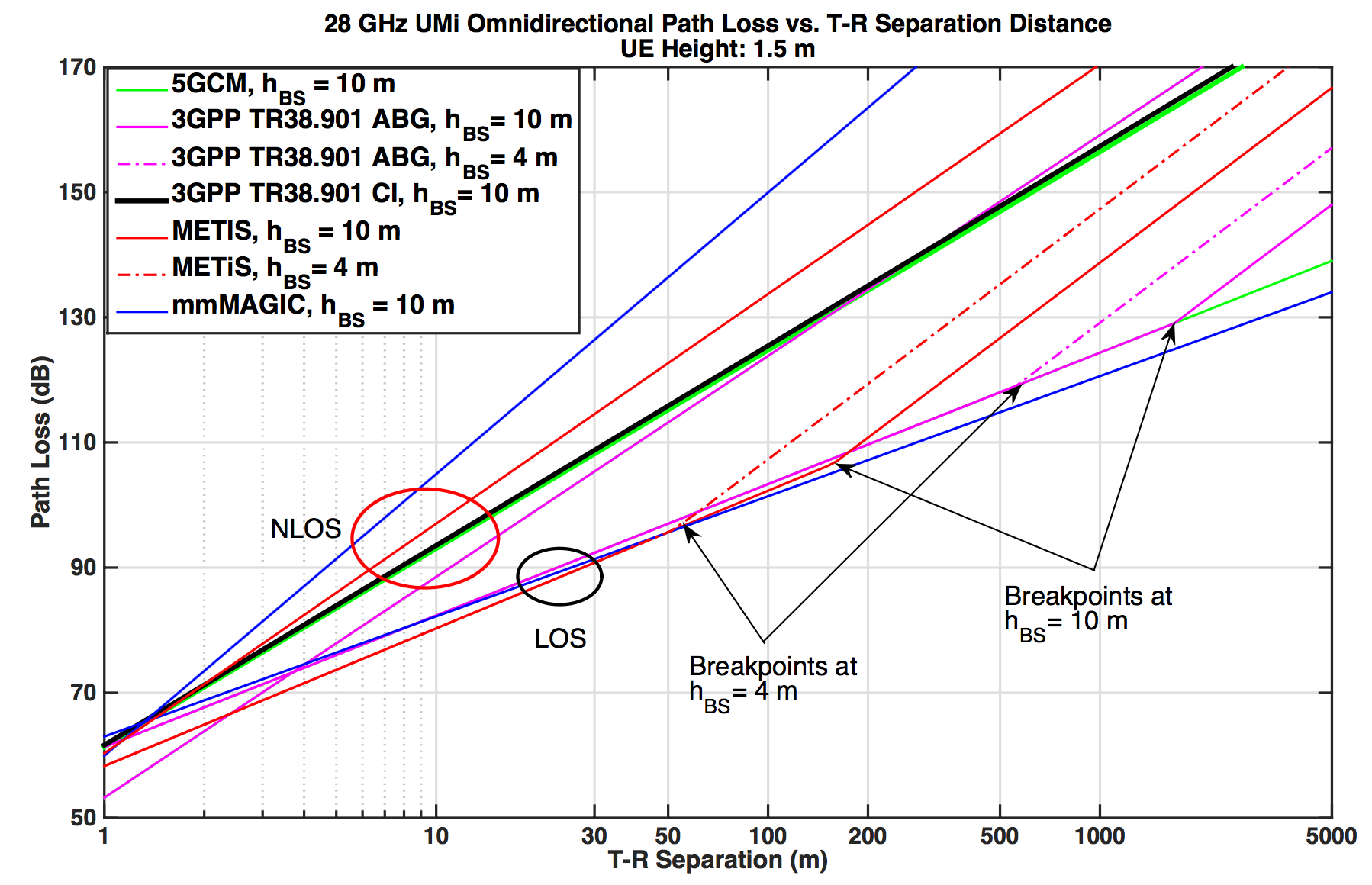}
    \caption{$PL$ vs. T-R distance comparison among four different path loss models in UMi scenario.}
    \label{fig:PLUMi}
\end{figure}
        
\paragraph{3GPP TR 38.901}\label{umi3gpp}
Path loss models in \cite{3GPP2017} use 3D T-R separation distances $d_{3D}$ that account for the BS height ($h_{BS}$) and UE height ($h_{UE}$). The distribution of the shadow fading is log-normal, and the standard deviation for LOS is $\sigma_{SF} = 4.0\;\text{dB}$. The UMi path loss model for LOS is a breakpoint model. For $d_{2D}<d_{BP}'$, the model is essentially a CI model with $n = 2.1$ \cite{andersen1995,sun2016propagation,rappaport2015wideband,Sun16b,sun2015path}. The LOS breakpoint distance $d_{BP}'$ is a function of the carrier frequency, BS height, and the UE height \cite{haneda20165g,3GPP2017}:
\begin{equation}\label{equ:UMi3GPPbp}
\footnotesize
\begin{split}
d_{BP}' = 4h_{BS}'h_{UE}'f_c \times 10^9/c \\
h_{BS}' = h_{BS}-1.0 \;\text{m}, \\
h_{UE}' = h_{UE}- 1.0 \;\text{m}
\end{split}
\end{equation}
where $h_{BS}'$ and $h_{UE}'$ are the effective antenna heights at the BS and the UE, and $h_{BS}$ and $h_{UE}$ are the actual antenna heights, respectively. The breakpoint distance in an urban environment \cite{bullington1947radio} is where the PLE transitions from free space ($n = 2$) to the asymptotic two-ray ground bounce model of $n = 4$ \cite{feuerstein1994path,JSAC}. At mmWave frequencies, the use of a breakpoint is controversial as it has not been reported in measurement, but some ray tracing simulations predict that it will occur \cite{hur2016proposal}. Since the UMi cells radius is typically 500 m or less, the use of a breakpoint and the height factors in (\ref{equ:UMi3GPPbp}) are not necessary (the breakpoint distance is larger than 500 m even with the smallest possible breakpoint distance when $h_{BS} = 4$ m and $h_{UE} = 1.5$ m as shown in Fig. \ref{fig:PLUMi}). The CI model provides a similar prediction of the path loss with a much simpler equation (\ref{equ:CI}) \cite{sun2015path}.

In the NLOS scenarios, the UMi-NLOS model uses the ABG model form \cite{hata1980}, with a frequency-dependent term that indicates path loss increases with frequency and also has an additional height correction term for the UE. Furthermore, a mathematical patch to correct model deficiencies is used to set a lower bound for the NLOS model as the LOS path loss. The shadow fading standard deviation for UMi NLOS is $\sigma_{SF} = 7.82 \;\text{dB}$ \cite{haneda20165g, piersanti2012millimeter,maccartney2013path}. The physically-based CI model is also provided as an optional NLOS path loss model for 3GPP TR 38.901 with parameter values given in Table~\ref{tbl:UMiPL}. 

\paragraph{METIS}\label{umimetis}
The path loss model for UMi in METIS~\cite{METIS2015} is a modified version of the ITU-R UMi path loss model \cite{ITU-RM.2135} and is claimed to be valid for frequencies from 0.8 to 60 GHz (see Table~\ref{tbl:UMiPL}). Some METIS models include breakpoints based on sub-6 GHz work (see Fig. \ref{fig:PLUMi}), yet mmWave measurements to date do not show breakpoints to exist \cite{hur2016proposal,JSAC,METIS2015}. For LOS scenarios, a scaling factor is used, so that the breakpoint distance $d_{BP}$ (in meters) becomes:

\begin{equation}
\footnotesize
\label{equ:METiSBP}
d_{BP} = 0.87 \exp \left( -\frac{\log_{10}(f_c)}{0.65} \right)  \frac{4 (h_{BS}-1 \text{m})(h_{UE}-1 \text{m})}{\lambda} 
\end{equation}
and  the path loss formula for LOS is written as: 
\begin{equation}
\footnotesize
\begin{split}
\label{equ:METiSPL}
PL_{LOS}(d_1)~\text{[dB]} = 10n_1\log_{10}\left( d_1\right) + 28.0 + 20\log_{10}\left( f_c \right) + PL_{0}
\end{split}
\end{equation}
for $10 \text{ m} < d \leqslant d_{BP} $, where $PL_{0}$ is a path loss offset calculated by:
\begin{equation} \label{PLoffset}
\footnotesize
PL_{0}~\text{[dB]} = -1.38 \log_{10}\left( f_c \right) + 3.34
\end{equation}

Path loss after the breakpoint distance is:
\begin{equation}
\label{equ:METiSPL2}
\footnotesize
\begin{split}
PL_{LOS}(d_1)~\text{[dB]} = 10n_2\log_{10}\left( \frac{d_1}{d_{BP}} \right)  + PL_{LOS}(d_{BP})
\end{split}
\end{equation}
for $d_{BP} < d_1 < 500 \text{ m}$ where \eqref{equ:METiSPL} and \eqref{equ:METiSPL2} represent path loss before and after the breakpoint, respectively. The last term $PL(d_{BP})$ in \eqref{equ:METiSPL2} is derived from (\ref{equ:METiSPL}) by substituting $d_1$ with $d_{BP}$ to calculate path loss at the breakpoint distance \cite{METIS2015}.

The UMi NLOS path loss model in METIS is adopted from the 3GPP TR 36.873 \cite{3GPP2014,METIS2015} sub-6 GHz model for 4G LTE and is calculated as:

\begin{equation}
\label{equ:METiSPLn}
\footnotesize
\begin{split}
&PL = \max{ \left( PL_{LOS}(d_{3D}), PL_{NLOS}(d_{3D}) \right) }\\
&PL_{NLOS}=36.7\log_{10}(d_{3D}) +23.15 +26\log_{10}(f_c) -0.3(h_{UE})
\end{split}
\end{equation}
where $f_c$ is in GHz, $10 \text{ m} < d_{3D} < 2000 \text{ m}$, and $1.5 \text{ m} \leq h_{UE} \leq 22.5 \text{ m}$.

\paragraph{mmMAGIC}
The mmMAGIC project \cite{mmMAGIC} adopted the ABG path loss model for UMi, similar to that from 5GCM \cite{5GCM} but with different parameter values (see Table~\ref{tbl:UMiPL}). Comparisons among the different UMi large-scale path loss models described here are provided in Fig. \ref{fig:PLUMi}.

\subsubsection{UMa Large-Scale Path Loss}
\paragraph{3GPP TR 38.901}
The 3GPP TR 38.901 \cite{3GPP2017} UMa LOS path loss model is adopted from 3GPP TR 36.873 (below 6 GHz Release 12 for LTE) \cite{3GPP2014} and TR 38.900 \cite{3GPP2016,3GPPTDOC}. For the UMa NLOS scenario, an ABG model and an optional CI model are provided (see Table~\ref{tbl:UMaPL} for parameters). With respect to the UMa LOS model, 3GPP TR 38.901 inexplicably discards the TR 38.900 \cite{3GPP2016} model and reverts back to TR 36.873 which is defined only for below 6 GHz \cite{3GPP2014} while also omitting the InH shopping mall scenario used in TR 38.900. TR 38.901 models omnidirectional path loss from 0.5-100 GHz, but lacks measurement validation in some cases.
  
\paragraph{5GCM}
There are three UMa path loss models used in \cite{5GCM}: CI, CIF, and ABG \cite{sun2015path,Sun16b}. The PLEs of the CI/CIF models for UMa are somewhat lower than for the UMi models indicating less loss over distance, which makes sense intuitively since a larger BS height implies that fewer obstructions are encountered than in the UMi scenario	 \cite{thomas2016prediction}. 

\paragraph{METIS}
METIS adopted the sub-6 GHz 3GPP TR 36.873 \cite{3GPP2014} 3D UMa model that was published in 2014 for LTE, see Table~\ref{tbl:UMaPL}.

\subsubsection{InH Large-Scale Path Loss}
\begin{table*}[!ht]\footnotesize
\renewcommand{\arraystretch}{1.1}
\centering
\caption{5GCM omnidirectional path loss models in the InH scenario} \label{tbl:InH5GCMPL}
\newcommand{\tabincell}[2]{\begin{tabular}{@{}#1@{}}#2\end{tabular}}
\begin{tabular}{|c|c|c|c|}
\hline
\tabincell{c}{ }&\tabincell{c}{\textbf{PL [dB], $f_c$ is in GHz, $d$ is in meters}}&\tabincell{c}{\textbf{Shadow fading} \\\textbf{std [dB]} }&\tabincell{c}{\textbf{Applicability range} \\ \textbf{and Parameters}}\\
\hline
\tabincell{c}{5GCM InH \\ Indoor-Office\\ LOS}&\tabincell{c}{\textbf{CI model with 1 m reference distance:}\\ $PL = 32.4 + 17.3 \log_{10}(d_{3D}) + 20 \log_{10}(f_c)$  }&\tabincell{c}{$\sigma_{SF} = 3.02$}&\tabincell{c}{$6 < f_c < 100 \;\text{GHz}$}\\
\hline
\tabincell{c}{5GCM InH \\ Indoor-Office\\ NLOS\\ single slope (FFS)}&\tabincell{c}{\textbf{CIF model:}\\ $PL = 32.4 + 31.9 (1+ 0.06(\frac{f_c-24.2 }{24.2}))\log_{10}(d_{3D}) + 20 \log_{10}(f_c)$\\ \textbf{ABG model:}\\ $PL = 38.3\log_{10}(d_{3D}) + 17.30 + 24.9\log_{10}(f_c)$}&\tabincell{c}{ \\$\sigma^{CIF}_{SF} = 8.29$\\ \\ $\sigma^{ABG}_{SF} = 8.03$}&\tabincell{c}{$6 < f_c < 100 \;\text{GHz}$}\\
\hline
\tabincell{c}{5GCM InH \\ Indoor-Office\\ NLOS\\ dual slope}&\tabincell{c}{\textbf{Dual-Slope CIF model:}\\ $PL_{Dual}^{CIF}(d) =\begin{cases}
FSPL(f_c , 1\;\text{m}) \\
+ 10n_1\left( 1+b_1\left( \frac{f_c-f_0}{f_0}\right)  \right)\log_{10}(d)  , & 1<d \leq d_{BP}\\
FSPL(f_c , 1\;\text{m}) \\
+ 10n_1\left( 1+b_1\left( \frac{f_c-f_0}{f_0}\right)  \right)\log_{10}(d_{BP})\\
+ 10n_2\left( 1+ b_2\left( \frac{f_c-f_0}{f_0} \right) \right) \log_{10}(\frac{d}{d_{BP}}), & d>d_{BP}
\end{cases}$\\ \textbf{Dual-Slope ABG model:}\\ $PL_{Dual}^{ABG}(d) =\begin{cases}
\alpha_1\cdot 10\log_{10} (d) + \beta_1 \\
+ \gamma \cdot 10\log_{10}(f_c) , & 1<d \leq d_{BP}\\
\alpha_1\cdot 10\log_{10} (d_{BP}) + \beta_1\\
 + \gamma \cdot 10\log_{10}(f_c)\\
+\alpha_2 \cdot 10 \log_{10}(\frac{d}{d_{BP}}), & d>d_{BP}
\end{cases}$}&\tabincell{c}{ \\$\sigma_{SF}^{CIF} = 7.65$\\ \\ $\sigma_{SF}^{ABG} = 7.78$}&\tabincell{c}{$6 < f_c < 100 \;\text{GHz}$\\ \textbf{Dual-Slope CIF model:}\\$n_1 = 2.51, b=0.06$\\ $f_0 = 24.1 \;\text{GHz}, n_2 =4.25 $\\ $b_2 =0.04 , d_{BP} = 7.8 \:\text{m}$ \\ \textbf{Dual-Slope ABG model:}\\ $\alpha_1 =1.7, \beta_1 = 33.0$\\ $\gamma = 2.49, d_{BP} = 6.9 \;\text{m}$\\ $\alpha_2 = 4.17$  }\\
\hline
\tabincell{c}{5GCM InH \\ Shopping-Mall\\ LOS}&\tabincell{c}{\textbf{CI model with 1 m reference distance:}\\ $PL = 32.4 + 17.3 \log_{10}(d_{3D}) + 20 \log_{10}(f_c)$  }&\tabincell{c}{$\sigma_{SF} = 2.01$}&\tabincell{c}{$6 < f_c < 100 \;\text{GHz}$}\\
\hline
\tabincell{c}{5GCM InH \\ Shopping-Mall\\ NLOS\\ single slope (FFS)}&\tabincell{c}{\textbf{CIF model:}\\ $PL = 32.4 + 25.9 (1+ 0.01(\frac{f_c-39.5 }{39.5 }))\log_{10}(d_{3D}) + 20 \log_{10}(f_c)$\\ \textbf{ABG model:}\\ $PL = 32.1\log_{10}(d_{3D}) + 18.09 + 22.4\log_{10}(f_c)$}&\tabincell{c}{ \\$\sigma_{SF}^{CIF} = 7.40$\\ \\ $\sigma_{SF}^{ABG} = 6.97$}&\tabincell{c}{$6 < f_c < 100 \;\text{GHz}$}\\
\hline
\tabincell{c}{5GCM InH \\ Shopping-Mall\\ NLOS\\ dual slope}&\tabincell{c}{\textbf{Dual-Slope CIF model:}\\ $PL_{Dual}^{CIF}(d) =\begin{cases}
FSPL(f_c , 1\;\text{m}) \\
+ 10n_1\left( 1+b_1\left( \frac{f_c-f_0}{f_0}\right)  \right)\log_{10}(d )  , & 1<d \leq d_{BP}\\
FSPL(f_c , 1\;\text{m}) \\
+ 10n_1\left( 1+b_1\left( \frac{f_c-f_0}{f_0}\right)  \right)\log_{10}(d_{BP})\\
+ 10n_2\left( 1+ b_2\left( \frac{f_c-f_0}{f_0} \right) \right) \log_{10}(\frac{d}{d_{BP}}), & d>d_{BP}
\end{cases}$\\ \textbf{Dual-Slope ABG model:}\\ $PL_{Dual}^{ABG}(d) =\begin{cases}
\alpha_1\cdot 10\log_{10} (d) + \beta_1 \\
+ \gamma \cdot 10\log_{10}(f_c) , & 1<d \leq d_{BP}\\
\alpha_1\cdot 10\log_{10} (d_{BP}) + \beta_1\\
 + \gamma \cdot 10\log_{10}(f_c)\\
+\alpha_2 \cdot 10 \log_{10}(\frac{d}{d_{BP}}), & d>d_{BP}
\end{cases}$}&\tabincell{c}{ \\$\sigma_{SF}^{CIF} = 6.26$\\ \\ $\sigma_{SF}^{ABG} = 6.36$}&\tabincell{c}{$6 < f_c < 100 \;\text{GHz}$\\ \textbf{Dual-Slope CIF model:}\\$n_1 = 2.43, b=-0.01$\\ $f_0 = 39.5 \;\text{GHz}, n_2 = 8.36 $\\ $b_2 =0.39 , d_{BP} = 110 \:\text{m}$ \\ \textbf{Dual-Slope ABG model:}\\ $\alpha_1 =2.9, \beta_1 = 22.17$\\ $\gamma = 2.24, d_{BP} = 147.0 \;\text{m}$\\ $\alpha_2 = 11.47$  }\\
\hline
\end{tabular}
\end{table*}

\begin{table*}[!ht]\footnotesize
\centering
\caption{Other omnidirectional path loss models in the InH scenario} \label{tbl:InHPL}
\newcommand{\tabincell}[2]{\begin{tabular}{@{}#1@{}}#2\end{tabular}}
\begin{tabular}{|c|c|c|c|}
\hline
\tabincell{c}{ }&\tabincell{c}{\textbf{PL [dB], $f_c$ is in GHz, $d$ is in meters}}&\tabincell{c}{\textbf{Shadow fading} \\\textbf{std [dB]} }&\tabincell{c}{\textbf{Applicability range} \\ \textbf{and Parameters}}\\
\hline
\multicolumn{4}{|c|}{\textbf{3GPP TR 38.901 V14.0.0 \cite{3GPP2017}} }\\
\hline
\tabincell{c}{3GPP TR 38.901\\ Indoor-Office LOS}&\tabincell{c}{$PL_{InH-LOS} = 32.4 + 17.3 \log_{10}(d_{3D}) + 20 \log_{10}(f_c)$}&\tabincell{c}{$\sigma_{SF} = 3.0$}&\tabincell{c}{$0.5 < f_c < 100 \;\text{GHz}$\\$1  < d_{3D} < 100 \;\text{m} $ }\\
\hline
\tabincell{c}{3GPP TR 38.901\\ Indoor-Office NLOS }&\tabincell{c}{$PL = \max{ \left( PL_{InH-LOS}(d_{3D}), PL_{InH-NLOS}(d_{3D}) \right) } $\\ $PL_{InH-NLOS} = 17.30+ 38.3\log_{10}(d_{3D}) + 24.9 \log_{10}(f_c)$\\ \textbf{Option: CI model with 1 m reference distance}\\$PL = 32.4 + 20\log_{10}(f_c)+31.9 \log_{10}(d_{3D})$}&\tabincell{c}{\\ $\sigma_{SF} = 8.03$\\ \\$\sigma_{SF} = 8.29$ }&\tabincell{c}{$0.5 < f_c < 100 \;\text{GHz}$\\$ 1 < d_{3D} < 86 \;\text{m}$ \\ \\  $ 1 < d_{3D} < 86 \;\text{m}$ }\\
\hline
\multicolumn{4}{|c|}{\textbf{METIS \cite{METIS2015}} }\\
\hline
\tabincell{c}{METIS \\ Shopping Mall LOS}&\tabincell{c}{$PL = 68.8 + 18.4 \log_{10}(d_{2D})$}&\tabincell{c}{$\sigma_{SF} = 2.0$}&\tabincell{c}{$ f_c =63 \;\text{GHz}$\\$1.5  < d_{2D} < 13.4 \;\text{m} $\\ $h_{BS}=h_{UE}=2 \;\text{m}$ }\\
\hline
\tabincell{c}{METIS \\ Shopping Mall NLOS}&\tabincell{c}{$PL = 94.3 + 3.59 \log_{10}(d_{2D})$}&\tabincell{c}{$\sigma_{SF} = 2.0$}&\tabincell{c}{$ f_c =63 \;\text{GHz}$\\$4  < d_{2D} < 16.1 \;\text{m} $\\ $h_{BS}=h_{UE}=2 \;\text{m}$ }\\
\hline
\multicolumn{4}{|c|}{\textbf{IEEE 802.11ad \cite{802.11ad}} }\\
\hline
\tabincell{c}{802.11ad \\ Indoor-Office LOS}&\tabincell{c}{$PL_{LOS}[dB] = 32.5 + 20 \log_{10}(f_c)+ 20 \log_{10}(d_{2D})$}&\tabincell{c}{$\sigma_{SF} $}&\tabincell{c}{$ 57 < f_c < 63 \;\text{GHz}$  }\\
\hline
\tabincell{c}{802.11ad \\ Indoor-Office NLOS}&\tabincell{c}{$PL_{NLOS}[dB] = 51.5 + 20 \log_{10}(f_c)+ 6 \log_{10}(d_{2D})$\\$PL_{NLOS}[dB] = 45.5 + 20 \log_{10}(f_c)+ 14 \log_{10}(d_{3D})$}&\tabincell{c}{$\sigma_{SF}^{STA-STA}=3.3 $ \\  $\sigma_{SF}^{STA-AP} = 3$}&\tabincell{c}{$ 57 < f_c < 63 \;\text{GHz}$}\\
\hline
\multicolumn{4}{|c|}{\textbf{mmMAGIC \cite{mmMAGIC}} }\\
\hline
\tabincell{c}{mmMAGIC InH \\ LOS}&\tabincell{c}{ $PL_{LOS} =  13.8 \log_{10}(d_{3D}) + 33.6 + 20.3 \log_{10}(f_c)$ }&\tabincell{c}{$\sigma_{SF} = 1.18$}&\tabincell{c}{$6 < f_c < 100 \;\text{GHz}$}\\
\hline
\tabincell{c}{mmMAGIC InH \\ NLOS}&\tabincell{c}{ $PL = \max{ \left( PL_{LOS}(d_{3D}), PL_{NLOS}(d_{3D}) \right) } $\\ $PL_{NLOS} =  36.9 \log_{10}(d_{3D}) + 15.2 + 26.8 \log_{10}(f_c)$ }&\tabincell{c}{$\sigma_{SF} = 8.03$}&\tabincell{c}{$6 < f_c < 100 \;\text{GHz}$}\\
\hline
\end{tabular}
\end{table*}

\paragraph{5GCM}
In the InH scenario, besides the CI, CIF, and ABG path loss models, dual-slope path loss models are proposed for different distance zones in the propagation environment and are provided in Table~\ref{tbl:InH5GCMPL}.
For NLOS, both the dual-slope ABG and dual-slope CIF models are considered for 5G performance evaluation, where they each require five modeling parameters to be optimized. Also, a single-slope CIF model that uses only two optimization parameters is considered for InH-Office~\cite{5GCM, maccartney2015indoor}. The dual-slope model may be best suited for InH-shopping mall or large indoor distances (greater than 50 m), although it is not clear from the data in \cite{5GCM} that the additional complexity is warranted when compared to the simple CIF model.

\paragraph{3GPP TR 38.901}
The path loss model for the InH-office LOS scenario in 3GPP TR 38.901 \cite{3GPP2017} is claimed to be valid up to 100 m and has the same form as the CI model in the UMi scenario. The only differences from UMi CI model are that the PLE in InH-office is slightly lower than that in the UMi street canyon due to more reflections and scattering in the indoor environment from walls and ceilings and waveguiding effects down hallways that increase received signal power \cite{maccartney2015indoor}.

The 3GPP TR 38.901 InH-office NLOS path loss model uses the ABG model form similar to its UMi NLOS path loss model, except that there is no height correction term, and the model requires a patch to ensure it is lower-bounded by the LOS path loss as follows:
\begin{equation}
\footnotesize
PL~\text{[dB]}= \max{ \left( PL_{InH-LOS}(d_{3D}), PL_{InH-NLOS}(d_{3D}) \right) }
\end{equation}
\begin{equation}
\footnotesize
\label{equ:3GPPInHN}
PL_{InH-NLOS}~\text{[dB]} = 17.30 + 38.3\log_{10}(d_{3D}) + 24.9\log_{10}(f_c)
\end{equation}


\paragraph{METIS}
In the latest METIS white paper \cite{METIS2015}, the WINNER II path loss model (similar in form to the ABG model) was adopted as the geometry-based stochastic model for short-range 60 GHz (61-65 GHz) links in indoor environments:
\begin{equation}\label{equ:METISInHPL}
\footnotesize
PL~\text{[dB]}= A \log_{10}(d) + B
\end{equation}
where $A$ and $B$ are curve-fit parameters without the use of Friis' equation \cite{friis1946note} (see Table~\ref{tbl:InHPL} for parameters).

\paragraph{mmMAGIC}
The InH channel model in mmMAGIC \cite{mmMAGIC} is adopted from an earlier version of 5GCM \cite{5GCM}, and has the same form as the ABG model. For Indoor-NLOS, the values of the path loss model parameters have been averaged from InH and InH-Shopping Mall. 

\paragraph{IEEE 802.11ad}
In the STA-STA  (STA signifies a station, the WiFi term for the UE) LOS scenario \cite{802.11ad}, path loss follows theoretical free space path loss in the CI model form via the Friis' free space transmission equation as given in Table~\ref{tbl:InHPL}. No shadowing term is provided in the LOS case, as instantaneous realizations are claimed to be close to the average path loss value over such wideband channel bandwidth. 

Experiments performed for NLOS situations resulted in path loss for STA-STA as a FI/AB model \cite{rappaport2015wideband} with the shadow fading standard deviation as $\sigma_{SF}=3.3$ dB. The 2D distance $d_{2D}$ is used for the STA-STA scenario, since it is considered that two stations are deemed to be at the same height above ground.

In the STA-AP (where the AP denotes access point, corresponding to a BS) scenario, the 3D separation distance $d_{3D}$ is used, and the LOS STA-AP path loss model is the same CI model as used in the STA-STA situation but no specific shadow fading term is given. The NLOS STA-AP model takes the same ABG form as that of STA-STA, but with $A_{NLOS} = 45.5$ dB and a shadow fading standard deviation $\sigma_{SF} = 3.0$ dB.  

\subsubsection{RMa Large-Scale Path Loss}

\paragraph{3GPP TR 38.901}
The 3GPP TR 38.901 RMa path loss model~\cite{3GPP2017} is mostly adopted from sub-6 GHz ITU-R M.2135~\cite{ITU-RM.2135} as described below, and claims validity up to 30 GHz, based on a single 24 GHz measurement campaign over short distances less than 500 m and without any goodness of fit indication~\cite{TDOC164975}. Work in \cite{Mac16c,JSAC} advocates a much more fundamental and accurate RMa model using the CIF model formulation in \eqref{equ:CIF}, where the frequency dependency of the PLE is replaced with a TX height dependency of the PLE, based on many propagation studies that showed UMa and RMa environment did not offer additional frequency dependency of the path loss over distance beyond the first meter of propagation \cite{Sun16b,Mac16c,JSAC,sun2016propagation}. 

\paragraph{ITU-R}
The ITU-R communication sector published guidelines for the evaluation of radio interface technologies for IMT-Advanced in ITU-R M.2135 which is valid for sub-6 GHz ~\cite{ITU-RM.2135}. The rural scenario is best described as having BS heights of 35 m or higher, generally much higher than surrounding buildings. The LOS path loss model has a controversial breakpoint distance \cite{JSAC} and a maximum 2D T-R separation distance of 10 km, while the NLOS path loss model has a maximum 2D T-R separation distance of 5 km with no breakpoint distance. Initial antenna height default values are provided in Table ~\ref{tbl:ITURMAappRange}, with the following four correction factor parameters: street width $W$, building height $h$, BS height $h_{BS}$, and UE height $h_{UE}$ (all in meters).

The ITU-R RMa LOS path loss model is quite complex:
\begin{align}\label{eq:RMaLOS}
\footnotesize
\begin{split}
PL _1~\text{[dB]}& = 20\log(40\pi \cdot d_{3D} \cdot f_c /3)\\
&+\min(0.03h^{1.72},10)\log_{10}(d_{3D}) \\
&-\min(0.044h^{1.72},14.77)+0.002\log_{10}(h)d_{3D}\\
PL_2~\text{[dB]} & = PL_1 (d_{BP})+40\log_{10}(d_{3D}/d_{BP})
\end{split}
\end{align}
where the breakpoint distance $d_{BP}$ is: 
\begin{equation}\label{eq:dbp}
\footnotesize
d_{BP} =  2\pi \cdot h_{BS} \cdot h_{UE} \cdot f_c/c
\end{equation}

It is must be noted that the model reverts to a single-slope model at 9.1 GHz or above, since the breakpoint distance exceeds 10 km (the outer limit of model applicability), thus making the LOS model mathematically inconsistent for mmWave frequencies above 9.1 GHz~\cite{Mac16c,JSAC}. 

The NLOS RMa path loss model in~\eqref{eq:RMaNLOS} is adopted from ITU-R M.2135 and has nine empirical coefficients for various building height and street width parameters~\cite{3GPP2017,ITU-RM.2135}:
\begin{align}\label{eq:RMaNLOS}
\footnotesize
\begin{split}
PL~\text{[dB]}& = \max(PL_{RMa-LOS},PL_{RMa-NLOS})\\
PL &_{RMa-NLOS}~\text{[dB]}= 161.04-7.1\log_{10}(W)+7.5\log_{10}(h)\\
&-(24.37-3.7(h/h_{BS})^2)\log_{10}(h_{BS})\\
&+(43.42-3.1\log_{10}(h_{BS}))(\log_{10}(d_{3D})-3)\\
&+20\log_{10}(f_c)-(3.2(\log_{10}(11.75h_{UE}))^2-4.97)
\end{split}
\end{align}

The ITU-R RMa NLOS path loss model from which the 3GPP TR38.901 model is adopted is only specified for frequencies up to 6 GHz and has not been validated in the literature for mmWave frequencies. The ITU-R RMa models were not developed using rural scenarios \cite{Mac16c,JSAC}, but instead were derived from measurements in downtown Tokyo, making them ill-suited for the RMa case.

\begin{table}
\renewcommand{\arraystretch}{1.1}
    \centering
    \caption{ITU-R M.2135/3GPP RMa path loss model default values and applicability ranges~\cite{ITU-RM.2135,3GPP2017}.}\label{tbl:ITURMAappRange}
    \scalebox{1}{
        \begin{tabu}{|l|}\hline
            \textbf{RMa LOS Default Values Applicability Range} \\ \specialrule{1.5pt}{0pt}{0pt}
            10 m $<d_{2D}<d_{BP}$, \\
            $d_{BP} < d_{2D} <10\:000$ m,\\
            $h_{BS} = 35$ m, $h_{UE}=1.5$ m, $W=20$ m, $h=5$ m\\
            Applicability ranges: 5 m $<h<50$ m; 5 m $<W<50$ m; \\
            10 m $<h_{BS}<150$ m; 1 m $<h_{UE}<10$ m \\ \hline
            \textbf{RMa NLOS Default Values Applicability Range} \\ \specialrule{1.5pt}{0pt}{0pt}
            10 m $<d_{2D}<5\:000$ m, \\
            
            $h_{BS} = 35$ m, $h_{UE}=1.5$ m, $W=20$ m, $h=5$ m\\
            Applicability ranges: 5 m $<h<50$ m; 5 m $<W<50$ m; \\
            10 m $<h_{BS}<150$ m; 1 m $<h_{UE}<10$ m \\ \hline
    \end{tabu}}
\end{table}

\paragraph{NYU RMa model}
NYU proposed empirically-based CIH RMa path loss models for LOS ($PL_{LOS}^{CIH-RMa}$) and NLOS ($PL_{NLOS}^{CIH-RMa}$) from extensive simulations and 73 GHz field data \cite{JSAC}:

\begin{align}
\footnotesize
\begin{split}
PL&_{LOS}^{CIH-RMa}(f_c,d,h_{BS})~\text{[dB]}=32.4+20\log_{10}(f_c)\\
&+23.1\left( 1-0.03\left(  \dfrac{h_{BS}-35}{35}\right)  \right) \log_{10}(d)+\chi_{\sigma_{LOS}}\\
\end{split}
\end{align}
where~$d \geq 1$~\text{m}, $\sigma_{LOS}=1.7$~\text{dB}, and~$10 \text{m} \leq h_{BS} \leq 150~\text{m}$.
\begin{align}
\footnotesize
\begin{split}
PL&_{NLOS}^{CIH-RMa}(f_c,d,h_{BS})~\text{[dB]}=32.4+20\log_{10}(f_c)\\
&+30.7\left( 1-0.049\left(  \dfrac{h_{BS}-35}{35}\right)  \right) \log_{10}(d)+\chi_{\sigma_{NLOS}}
\end{split}
\end{align}
where $d \geq 1 \text{m}$, $\sigma_{LOS}=6.7~\text{dB}$, and $10 \text{m} \leq h_{BS} \leq 150~\text{m}$.

\subsection{O2I Penetration Loss} 
\subsubsection{3GPP TR 38.901}
The overall large-scale path loss models may also account for penetration loss into a building and subsequent path loss inside the building. The O2I path loss model taking account of the building penetration loss according to 3GPP TR 38.901 \cite{3GPP2017} has the following form:
\begin{equation}
\label{equ:O2I}
\footnotesize
PL~\text{[dB]}= PL_b + PL_{tw} + PL_{in} + N(0,\sigma_P^2)
\end{equation}
where $PL_b$ is the basic outdoor path loss, $PL_{tw}$ is the building penetration loss through the external wall, $PL_{in}$ is the indoor loss which depends on the depth into the building, and $\sigma_P$ is the standard deviation for the penetration loss. The building penetration loss $PL_{tw}$ can be modeled as:
\begin{equation}
\footnotesize
PL_{tw}~\text{[dB]} = PL_{npi} - 10\log_{10}\sum_{i=1}^{N}\left( p_i \times 10^{\frac{L_{\text{material}_i}}{-10}} \right) 
\end{equation}
where $PL_{npi}$ is an additional loss which is added to the external wall loss to account for non-perpendicular incidence, $L_{\text{material}_i} =a_{\text{material}_i} +b_{\text{material}_i}\cdot f_c $ is the penetration loss of material $i$, $f_c$ is the frequency in GHz, $p_i$ is the proportion of $i$-th materials, where $\sum p_i =1$, and $N$ is the number of materials. Penetration loss of several materials and the O2I penetration loss models are given in Table \ref{tbl:O2Imat}.

\begin{table}[h]\footnotesize
\renewcommand{\arraystretch}{1.1}
\centering
\caption{ O2I penetration loss of different materials \cite{3GPP2017}}\label{tbl:O2Imat}
\newcommand{\tabincell}[2]{\begin{tabular}{@{}#1@{}}#2\end{tabular}}
\begin{tabular}{|c|c|}
\hline
\tabincell{c}{\textbf{Material}}&\tabincell{c}{\textbf{Penetration loss [dB], $f_c$ is in GHz}}\\
\hline
\tabincell{c}{Standard multi-pane glass}&\tabincell{c}{$L_{\text{glass}}= 2+ 0.2 \cdot f_c$}\\
\hline
\tabincell{c}{IRR glass}&\tabincell{c}{$L_{\text{IRRglass}}= 23 + 0.3 \cdot f_c$}\\
\hline
\tabincell{c}{Concrete}&\tabincell{c}{$L_{\text{concrete}}= 5 + 4 \cdot f_c$}\\
\hline
\tabincell{c}{Wood}&\tabincell{c}{$L_{\text{wood}}= 4.85 + 0.12 \cdot f_c$}\\
\hline

\end{tabular}
\end{table}
Rough models are also provided to estimate the building penetration loss in Table~\ref{tab:O2Ipeneloss}. Both the low-loss and high-loss models are applicable to UMa and UMi-street canyon, while only the low-loss model is applicable to RMa. The O2I car penetration loss included in path loss is determined by:

\begin{equation}
\footnotesize
\label{equ:car}
PL~\text{[dB]}= PL_b + N(\mu,\sigma_P^2)
\end{equation} 
where $PL_b$ is the basic outdoor path loss, and for most cases, $\mu=9 \;\text{dB}$ and $\sigma_P = 5 \;\text{dB}$. An optional $\mu= 20 \;\text{dB}$ is provided for metalized car windows for frequencies ranging from 0.6 to 60 GHz \cite{3GPP2017}. 

\begin{table*}[!ht]\footnotesize
\renewcommand{\arraystretch}{1.1}
\centering
\caption{O2I penetration loss parameters \cite{3GPP2017,5GCM}}\label{tab:O2Ipeneloss}
\newcommand{\tabincell}[2]{\begin{tabular}{@{}#1@{}}#2\end{tabular}}
\begin{tabular}{|c|c|c|c|}
\hline
\tabincell{c}{ }&\tabincell{c}{\textbf{Path loss through external wall:}\\$PL_{tw}$ \textbf{[dB], $f_c$ is in GHz}}&\tabincell{c}{\textbf{Indoor loss:} \\ $PL_{in}$ \textbf{[dB], d is in meters}}&\tabincell{c}{\textbf{Standard deviation:}\\$\sigma_P$  \textbf{[dB]}}\\
\hline
\tabincell{c}{3GPP TR 38.901  Low-loss model \cite{3GPP2017}}&\tabincell{c}{$5-10\log_{10}(0.3\cdot 10^{-L_{glass}/10}+0.7\cdot 10^{-L_{concrete}/10})$}&\tabincell{c}{$0.5d_{2D-in}$}&\tabincell{c}{4.4}\\
\hline
\tabincell{c}{3GPP TR 38.901 High-loss model \cite{3GPP2017}}&\tabincell{c}{$5-10\log_{10}(0.7\cdot 10^{-L_{IRRglass}/10}+0.3\cdot 10^{-L_{concrete}/10})$}&\tabincell{c}{$0.5d_{2D-in}$}&\tabincell{c}{6.5}\\
\hline
\tabincell{c}{5GCM Low-loss model \cite{5GCM,rappaport20175g}}&\tabincell{c}{$ 10 \log_{10}(5 + 0.03 \cdot f_c^2)$}&\tabincell{c}{Not Specified}&\tabincell{c}{4.0}\\
\hline
\tabincell{c}{5GCM High-loss model \cite{5GCM,rappaport20175g}}&\tabincell{c}{$ 10 \log_{10}(10 + 5 \cdot f_c^2)$}&\tabincell{c}{Not Specified}&\tabincell{c}{6.0}\\
\hline
\end{tabular}
\end{table*}

\subsubsection{5GCM}
The 5GCM adopted the building penetration loss model of 3GPP TR 36.873 which is based on legacy measurements below 6 GHz \cite{3GPP2014}. Several different frequency-dependent models were also proposed in \cite{haneda20165g,5GCM}. In \cite{semaan2014outdoor}, a detailed description of external wall penetration loss using a composite approach is provided. The difference of the building penetration loss model between 5GCM and 3GPP TR 38.901 is that the standard deviation is tentatively selected from the measurement data \cite{haneda20165g, semaan2014outdoor}. A very simple parabolic model with a good fit for predicting building penetration loss (BPL) of either high loss or low loss buildings was provided in \cite{haneda20165g,rappaport20175g} as:
\begin{equation}\label{equ:nyuO2I}
\footnotesize
\begin{split}
BPL~\text{[dB]} = 10 \log_{10}(A + B \cdot f_c^2)
\end{split}
\end{equation}
where $f_c$ is in GHz, $A=5$, and $B=0.03$ for low loss buildings and $A=10$ and $B= 5$ for high loss buildings.

\subsubsection{mmMAGIC}
The O2I penetration loss model in mmMAGIC has the form of:
\begin{equation}\label{equ:mmO2I}
\footnotesize
\begin{split}
O2I~\text{[dB]} = B_{O2I} + C_{O2I} \cdot \log_{10}\left( f_c \right) \approx 8.5 + 11.2\cdot \log_{10} \left( f_c \right) 
\end{split}
\end{equation}
The advantage of this form is that the coefficients $B_{O2I}$ and $C_{O2I}$ can be added to the existing coefficients in the path loss model of mmMAGIC. A frequency-dependent shadow fading between 8 and 10 dB for the UMi-O2I scenario is presented in \cite{mmMAGIC}:
\begin{equation}\label{equ:mmO2Isf}
\footnotesize
\begin{split}
\Sigma_{SF}~\text{[dB]} = \sigma_{SF} + \delta_{SF}\cdot \log_{10}\left( f_c \right) \approx 5.7 + 2.3\cdot \log_{10} \left( f_c \right) 
\end{split}
\end{equation}
\subsection{Spatial consistency}
\label{spatial}
Many previous channel models were ``drop-based'', where a UE is placed at a random location, random channel parameters (conditioned on this location) are assigned, performance is computed (possibly when moving over a short distance, up to 40 wavelengths), and then a different location is chosen at random. This approach is useful for statistical or monte-carlo performance analysis, but does not provide spatial consistency, i.e., two UEs that are dropped at nearly identical T-R separation distances might experience completely different channels from a system simulator. The importance of spatial consistency is dependent upon the site-specific propagation in a particular location as shown in \cite{rumney2016testing2,rap2016ap}. Channel models of 5GCM \cite{5GCM}, 3GPP TR 38.901 \cite{3GPP2017}, METIS \cite{METIS2015} and MiWEBA \cite{MiWEBA} provide new approaches for modeling of trajectories to retain spatial consistency.

In 5GCM and 3GPP, both the LOS/NLOS state and the shadowing states are generated on a coarse grid, and spatially filtered. This resulting ``map'' of LOS states and shadowing attenuations are then used for the trajectories of all UEs during the simulation process. For the implementation of the LOS state filtering, different methods are proposed \cite{5GCM,3GPP2017}, but the effect is essentially the same. We note that 5GCM and 3GPP also introduce additional procedures to ensure spatial consistencies of the delay and angles, but those considerations are beyond the scope of this paper. The map-based models of METIS \cite{METIS2015} and MiWEBA \cite{MiWEBA} inherently provide spatial consistency, as the dominant paths for close-by locations are identical, and their effect is computed deterministically. Generally speaking, spatial consistency is easier to implement in geometry-based models (such as semi-deterministic and geometric-based stochastic channel models) than in tapped-delay line models such as 3GPP. Work in \cite{rumney2016testing2,mmMAGIC,5GCM,samimi20163,rap2016ap} shows that the degree of spatial consistency can vary widely at mmWave frequencies.


\section{Conclusion}~\label{sec:conc}
Often times, standard bodies have additional reasons to adopt particular modeling formulations, beyond physical laws or the fitting of data to observed channel characteristics. Motivations often include ensuring simulations work for legacy software at lower frequencies, or the desire to rapidly converge while preserving legacy approaches (see \cite{sun2017a,Sun16b,maccartney2017study,JSAC} for example). Channel modeling for 5G is an on-going process and early results show significant capacity differences arise from different models \cite{rappaport20175g,sun2017a,rappaport2017VTC}. Futher work is needed to bolster and validate the early channel models. Many new mmWave channel simulators (e.g., NYUSIM, QuaDRiGa) have been developed and are being used by researchers to evaluate the performance of communication systems and to simulate channel characteristics when designing air interfaces or new wireless technologies across the network stack \cite{sun2017a,rajendran2011concepts,jaeckel2014quadriga,NYUSIM}. 

This paper has provided a comprehensive overview of emerging 5G mmWave wireless system concepts, and has provided a compilation of important mmWave radio propagation models developed throughout the world to date. The paper demonstrates early standards work and illustrates the various models obtained by several independent groups based on extensive measurements and ray tracing methods at mmWave frequency bands in various scenarios.

The development of proper propagation models is vital, not only for the long-term development of future mmWave wireless systems but also for fundamental understanding by future engineers and students who will learn about and improve the nascent mmWave mobile industry that is just now being developed. Various companies have started 5G field trials, and some of them have achieved 20 Gbps date rates \cite{HUAWEI,ERICSSON}. The fundamental information on path loss and shadowing surveyed in this paper is a prerequisite for moving further along the road to 5G at the unprecedented mmWave frequency bands.

\bibliographystyle{IEEEtran}
\bibliography{5Gpaper_draft_v_1_0}

\begin{thebibliography}{100}
\providecommand{\url}[1]{#1}
\csname url@samestyle\endcsname
\providecommand{\newblock}{\relax}
\providecommand{\bibinfo}[2]{#2}
\providecommand{\BIBentrySTDinterwordspacing}{\spaceskip=0pt\relax}
\providecommand{\BIBentryALTinterwordstretchfactor}{4}
\providecommand{\BIBentryALTinterwordspacing}{\spaceskip=\fontdimen2\font plus
\BIBentryALTinterwordstretchfactor\fontdimen3\font minus
  \fontdimen4\font\relax}
\providecommand{\BIBforeignlanguage}[2]{{%
\expandafter\ifx\csname l@#1\endcsname\relax
\typeout{** WARNING: IEEEtran.bst: No hyphenation pattern has been}%
\typeout{** loaded for the language `#1'. Using the pattern for}%
\typeout{** the default language instead.}%
\else
\language=\csname l@#1\endcsname
\fi
#2}}
\providecommand{\BIBdecl}{\relax}
\BIBdecl

\bibitem{gubbi2013internet}
J.~Gubbi \emph{et~al.}, ``{Internet of Things (IoT)}: A vision, architectural
  elements, and future directions,'' \emph{Future Generation Computer Systems},
  vol.~29, no.~7, pp. 1645--1660, Sept. 2013.

\bibitem{rapfcc16}
T.~S. Rappaport, ``{Spectrum Frontiers: The New World of Millimeter-Wave Mobile
  Communication},'' \emph{Invited keynote presentation, The Federal
  Communications Commission (FCC) Headquarters}, Mar.\ 10 2016.

\bibitem{rappaport2013millimeter}
T.~S. Rappaport \emph{et~al.}, ``{Millimeter Wave Mobile Communications for
  {5G} Cellular: It Will Work!}'' \emph{IEEE Access}, vol.~1, pp. 335--349, May
  2013.

\bibitem{rangan2014millimeter}
S.~Rangan, T.~S. Rappaport, and E.~Erkip, ``Millimeter-wave cellular wireless
  networks: Potentials and challenges,'' \emph{Proceedings of the IEEE}, vol.
  102, no.~3, pp. 366--385, Mar. 2014.

\bibitem{ghosh2014mmwave}
A.~Ghosh \emph{et~al.}, ``Millimeter-wave enhanced local area systems: A
  high-data-rate approach for future wireless networks,'' \emph{IEEE Journal on
  Selected Areas in Communications}, vol.~32, no.~6, pp. 1152--1163, June 2014.

\bibitem{roh2014millimeter}
W.~Roh \emph{et~al.}, ``Millimeter-wave beamforming as an enabling technology
  for {5G} cellular communications: theoretical feasibility and prototype
  results,'' \emph{IEEE Communications Magazine}, vol.~52, no.~2, pp. 106--113,
  Feb. 2014.

\bibitem{FCC16-89}
\BIBentryALTinterwordspacing
{Federal Communications Commission}, ``{Spectrum Frontiers Report and Order and
  Further Notice of Proposed Rulemaking: FCC16-89},'' July 2016. [Online].
  Available:
  \url{https://apps.fcc.gov/edocs\_public/attachmatch/FCC-16-89A1\_Rcd.pdf}
\BIBentrySTDinterwordspacing

\bibitem{singh2015tractable}
S.~Singh \emph{et~al.}, ``Tractable model for rate in self-backhauled
  millimeter wave cellular networks,'' \emph{IEEE Journal on Selected Areas in
  Communications}, vol.~33, no.~10, pp. 2196--2211, Oct. 2015.

\bibitem{sundaresan2016fluidnet}
K.~Sundaresan \emph{et~al.}, ``Fluidnet: a flexible cloud-based radio access
  network for small cells,'' \emph{IEEE/ACM Transactions on Networking},
  vol.~24, no.~2, pp. 915--928, Apr. 2016.

\bibitem{banelli2014modulation}
P.~Banelli \emph{et~al.}, ``Modulation formats and waveforms for {5G} networks:
  Who will be the heir of {OFDM}?: An overview of alternative modulation
  schemes for improved spectral efficiency,'' \emph{IEEE Signal Processing
  Magazine}, vol.~31, no.~6, pp. 80--93, Nov. 2014.

\bibitem{michailow2014generalized}
N.~Michailow \emph{et~al.}, ``Generalized frequency division multiplexing for
  5th generation cellular networks,'' \emph{IEEE Transactions on
  Communications}, vol.~62, no.~9, pp. 3045--3061, Sept. 2014.

\bibitem{5GCM}
\BIBentryALTinterwordspacing
5GCM, ``{5G} {Channel} {Model} for bands up to 100 {GHz},'' Tech. Rep., Oct.
  2016. [Online]. Available: \url{http://www.5gworkshops.com/5GCM.html}
\BIBentrySTDinterwordspacing

\bibitem{haneda2016indoor}
K.~Haneda \emph{et~al.}, ``Indoor {5G 3GPP}-like channel models for office and
  shopping mall environments,'' in \emph{2016 IEEE International Conference on
  Communications Workshops (ICC)}, May 2016, pp. 694--699.

\bibitem{deng201528}
S.~Deng, M.~K. Samimi, and T.~S. Rappaport, ``28 {GHz} and 73 {GHz}
  millimeter-wave indoor propagation measurements and path loss models,'' in
  \emph{IEEE International Conference on Communications Workshops (ICCW)}, June
  2015, pp. 1244--1250.

\bibitem{rappaport201573}
T.~S. Rappaport and S.~Deng, ``73 {GHz} wideband millimeter-wave foliage and
  ground reflection measurements and models,'' in \emph{2015 IEEE International
  Conference on Communication Workshop (ICCW)}, June 2015, pp. 1238--1243.

\bibitem{haneda20165g}
K.~Haneda \emph{et~al.}, ``{5G 3GPP}-like channel models for outdoor urban
  microcellular and macrocellular environments,'' in \emph{2016 IEEE 83rd
  Vehicular Technology Conference (VTC 2016-Spring)}, May 2016, pp. 1--7.

\bibitem{nie201372}
S.~Nie \emph{et~al.}, ``72 {GHz} millimeter wave indoor measurements for
  wireless and backhaul communications,'' in \emph{2013 IEEE 24th International
  Symposium on Personal Indoor and Mobile Radio Communications (PIMRC)}, Sept.
  2013, pp. 2429--2433.

\bibitem{haneda2016frequency}
K.~Haneda \emph{et~al.}, ``Frequency-agile pathloss models for urban street
  canyons,'' \emph{IEEE Transactions on Antennas and Propagation}, vol.~64,
  no.~5, pp. 1941--1951, May 2016.

\bibitem{JSAC}
G.~R. MacCartney and T.~S. Rappaport, ``Rural macrocell path loss models for
  millimeter wave wireless communications,'' \emph{IEEE Journal on Selected
  Areas in Communications}, vol.~35, no.~7, pp. 1663--1677, July 2017.

\bibitem{rappaport2015wideband}
T.~S. Rappaport \emph{et~al.}, ``Wideband millimeter-wave propagation
  measurements and channel models for future wireless communication system
  design,'' \emph{IEEE Transactions on Communications}, vol.~63, no.~9, pp.
  3029--3056, Sept. 2015.

\bibitem{maccartney2015indoor}
G.~R. {MacCartney, Jr.} \emph{et~al.}, ``Indoor office wideband millimeter-wave
  propagation measurements and models at 28 {GHz} and 73 {GHz} for ultra-dense
  {5G} wireless networks,'' \emph{IEEE Access}, pp. 2388--2424, Oct. 2015.

\bibitem{maccartney2013path}
------, ``Path loss models for {5G} millimeter wave propagation channels in
  urban microcells,'' in \emph{2013 IEEE Global Communications Conference
  (GLOBECOM)}, Dec. 2013, pp. 3948--3953.

\bibitem{samimi2015probabilistic}
M.~K. Samimi, T.~S. Rappaport, and G.~R. {MacCartney, Jr.}, ``Probabilistic
  omnidirectional path loss models for millimeter-wave outdoor
  communications,'' \emph{IEEE Wireless Communications Letters}, vol.~4, no.~4,
  pp. 357--360, Aug. 2015.

\bibitem{Mac16c}
G.~R. {MacCartney, Jr.} \emph{et~al.}, ``Millimeter wave wireless
  communications: New results for rural connectivity,'' in \emph{All Things
  Cellular'16: Workshop on All Things Cellular Proceedings, in conjunction with
  ACM MobiCom}, Oct. 2016, pp. 31--36.

\bibitem{sun2016millimeter}
S.~Sun, G.~R. {MacCartney, Jr.}, and T.~S. Rappaport, ``Millimeter-wave
  distance-dependent large-scale propagation measurements and path loss models
  for outdoor and indoor {5G} systems,'' in \emph{2016 IEEE 10th European
  Conference on Antennas and Propagation (EuCAP)}, Apr. 2016, pp. 1--5.

\bibitem{maccartney2015exploiting}
G.~R. {MacCartney, Jr.}, M.~K. Samimi, and T.~S. Rappaport, ``Exploiting
  directionality for millimeter-wave wireless system improvement,'' in
  \emph{2015 IEEE International Conference on Communications (ICC)}, June 2015,
  pp. 2416--2422.

\bibitem{thomas2016prediction}
T.~A. Thomas \emph{et~al.}, ``A prediction study of path loss models from
  2-73.5 {GHz} in an urban-macro environment,'' in \emph{2016 IEEE 83rd
  Vehicular Technology Conference (VTC 2016-Spring)}, May 2016, pp. 1--5.

\bibitem{Sun16b}
S.~Sun \emph{et~al.}, ``Investigation of prediction accuracy, sensitivity, and
  parameter stability of large-scale propagation path loss models for {5G}
  wireless communications,'' \emph{IEEE Transactions on Vehicular Technology},
  vol.~65, no.~5, pp. 2843--2860, May 2016.

\bibitem{samimi20163}
M.~K. Samimi and T.~S. Rappaport, ``{3-D} millimeter-wave statistical channel
  model for {5G} wireless system design,'' \emph{IEEE Transactions on Microwave
  Theory and Techniques}, vol.~64, no.~7, pp. 2207--2225, July 2016.

\bibitem{hur2014syn}
S.~Hur \emph{et~al.}, ``Synchronous channel sounder using horn antenna and
  indoor measurements on 28 {GHz},'' in \emph{2014 IEEE International Black Sea
  Conference on Communications and Networking (BlackSeaCom)}, May 2014, pp.
  83--87.

\bibitem{rappaport2013broadband}
T.~S. Rappaport \emph{et~al.}, ``Broadband millimeter-wave propagation
  measurements and models using adaptive-beam antennas for outdoor urban
  cellular communications,'' \emph{IEEE Transactions on Antennas and
  Propagation}, vol.~61, no.~4, pp. 1850--1859, Apr. 2013.

\bibitem{Koymen15a}
O.~H. Koymen \emph{et~al.}, ``{Indoor mm-Wave Channel Measurements: Comparative
  Study of 2.9 {GHz} and 29 {GHz}},'' in \emph{2015 IEEE Global
  Telecommunications Conference Workshops (Globecom Workshops)}, Dec. 2015, pp.
  1--6.

\bibitem{fettweis2014tactile}
G.~P. Fettweis, ``The tactile internet: applications and challenges,''
  \emph{IEEE Vehicular Technology Magazine}, vol.~9, no.~1, pp. 64--70, Mar.
  2014.

\bibitem{mecklenbrauker2011vehicular}
C.~F. Mecklenbrauker \emph{et~al.}, ``Vehicular channel characterization and
  its implications for wireless system design and performance,''
  \emph{Proceedings of the IEEE}, vol.~99, no.~7, pp. 1189--1212, July 2011.

\bibitem{gozalvez2012ieee}
J.~Goz{\'a}lvez, M.~Sepulcre, and R.~Bauza, ``{IEEE} 802.11 p vehicle to
  infrastructure communications in urban environments,'' \emph{IEEE
  Communications Magazine}, vol.~50, no.~5, pp. 176--183, May 2012.

\bibitem{bhushan2014network}
N.~Bhushan \emph{et~al.}, ``Network densification: the dominant theme for
  wireless evolution into {5G},'' \emph{IEEE Communications Magazine}, vol.~52,
  no.~2, pp. 82--89, Feb. 2014.

\bibitem{maeder2011challenge}
A.~Maeder \emph{et~al.}, ``The challenge of {M2M} communications for the
  cellular radio access network,'' in \emph{W{\"u}rzburg Workshop on IP: Joint
  ITG and Euro-NF Workshop” Visions of Future Generation
  Networks”(EuroView)}, Aug. 2011, pp. 1--2.

\bibitem{nikravesh2016depth}
A.~Nikravesh \emph{et~al.}, ``An in-depth understanding of multipath tcp on
  mobile devices: measurement and system design,'' in \emph{Proceedings of the
  22nd Annual International Conference on Mobile Computing and Networking},
  Oct. 2016, pp. 189--201.

\bibitem{deng2014wifi}
S.~Deng \emph{et~al.}, ``{WiFi}, {LTE}, or both?: Measuring multi-homed
  wireless internet performance,'' in \emph{Proceedings of the 2014 Conference
  on Internet Measurement Conference}, Nov. 2014, pp. 181--194.

\bibitem{andrews2014will}
J.~G. Andrews \emph{et~al.}, ``What will {5G} be?'' \emph{IEEE Journal on
  Selected Areas in Communications}, vol.~32, no.~6, pp. 1065--1082, June 2014.

\bibitem{backhaul2015}
E.~Bastug, M.~Bennis, and M.~Debbah, ``Living on the edge: The role of
  proactive caching in {5G} wireless networks,'' \emph{IEEE Communications
  Magazine}, vol.~52, no.~8, pp. 82--89, Aug 2014.

\bibitem{carapellese2014energy}
N.~Carapellese \emph{et~al.}, ``An energy consumption comparison of different
  mobile backhaul and fronthaul optical access architectures,'' in \emph{The
  European Conference on Optical Communication (ECOC)}, Sept. 2014, pp. 1--3.

\bibitem{hur2013millimeter}
S.~Hur \emph{et~al.}, ``Millimeter wave beamforming for wireless backhaul and
  access in small cell networks,'' \emph{IEEE Transactions on Communications},
  vol.~61, no.~10, pp. 4391--4403, Oct. 2013.

\bibitem{H2020}
\BIBentryALTinterwordspacing
{H2020 Project 5G-XHaul }, ``{Dynamically Reconfigurable Optical-Wireless
  Backhaul/Fronthaul with Cognitive Control Plane for Small Cells and
  Cloud-RANs},'' 2015. [Online]. Available:
  \url{http://www.5g-xhaul-project.eu/index.html}
\BIBentrySTDinterwordspacing

\bibitem{chandrasekhar2008femtocell}
V.~Chandrasekhar, J.~G. Andrews, and A.~Gatherer, ``Femtocell networks: a
  survey,'' \emph{IEEE Communications Magazine}, vol.~46, no.~9, pp. 59--67,
  Sept. 2008.

\bibitem{dohler2011phy}
M.~Dohler \emph{et~al.}, ``Is the {PHY} layer dead?'' \emph{IEEE Communications
  Magazine}, vol.~49, no.~4, pp. 159--165, Apr. 2011.

\bibitem{wang2014cellular}
C.-X. Wang \emph{et~al.}, ``Cellular architecture and key technologies for {5G}
  wireless communication networks,'' \emph{IEEE Communications Magazine},
  vol.~52, no.~2, pp. 122--130, Feb. 2014.

\bibitem{ge20145g}
X.~Ge \emph{et~al.}, ``{5G} wireless backhaul networks: challenges and research
  advances,'' \emph{IEEE Network}, vol.~28, no.~6, pp. 6--11, Nov. 2014.

\bibitem{george2014backhaul}
G.~R. MacCartney and T.~S. Rappaport, ``73 ghz millimeter wave propagation
  measurements for outdoor urban mobile and backhaul communications in new york
  city,'' in \emph{2014 IEEE International Conference on Communications (ICC)},
  June 2014, pp. 4862--4867.

\bibitem{murdock2014consumption}
J.~N. Murdock and T.~S. Rappaport, ``Consumption factor and power-efficiency
  factor: A theory for evaluating the energy efficiency of cascaded
  communication systems,'' \emph{IEEE Journal on Selected Areas in
  Communications}, vol.~32, no.~2, pp. 221--236, Feb. 2014.

\bibitem{haider2011spectral}
F.~Haider \emph{et~al.}, ``Spectral efficiency analysis of mobile femtocell
  based cellular systems,'' in \emph{2011 IEEE 13th International Conference on
  Communication Technology (ICCT)}, Sept. 2011, pp. 347--351.

\bibitem{hossain2014evolution}
E.~Hossain \emph{et~al.}, ``Evolution toward {5G} multi-tier cellular wireless
  networks: An interference management perspective,'' \emph{IEEE Wireless
  Communications}, vol.~21, no.~3, pp. 118--127, June 2014.

\bibitem{andrews2014an}
J.~G. Andrews \emph{et~al.}, ``An overview of load balancing in hetnets: old
  myths and open problems,'' \emph{IEEE Wireless Communications}, vol.~21,
  no.~2, pp. 18--25, Apr. 2014.

\bibitem{tehrani2014device}
M.~N. Tehrani, M.~Uysal, and H.~Yanikomeroglu, ``Device-to-device communication
  in {5G} cellular networks: challenges, solutions, and future directions,''
  \emph{IEEE Communications Magazine}, vol.~52, no.~5, pp. 86--92, May 2014.

\bibitem{yang2015software}
M.~Yang \emph{et~al.}, ``Software-defined and virtualized future mobile and
  wireless networks: A survey,'' \emph{Mobile Networks and Applications},
  vol.~20, no.~1, pp. 4--18, Sept. 2015.

\bibitem{agyapong2014design}
P.~K. Agyapong \emph{et~al.}, ``Design considerations for a 5g network
  architecture,'' \emph{IEEE Communications Magazine}, vol.~52, no.~11, pp.
  65--75, Nov. 2014.

\bibitem{van2008out}
J.~Van De~Beek and F.~Berggren, ``Out-of-band power suppression in {OFDM},''
  \emph{IEEE communications letters}, vol.~12, no.~9, pp. 609--611, Sept. 2008.

\bibitem{monk2016otfs}
A.~Monk \emph{et~al.}, ``{OTFS-Orthogonal Time Frequency Space},''
  \emph{Computing Research Repository (CoRR)}, vol. abs/1608.02993, Aug. 2016.

\bibitem{sun2014mimo}
S.~Sun \emph{et~al.}, ``{MIMO} for millimeter-wave wireless communications:
  beamforming, spatial multiplexing, or both?'' \emph{IEEE Communications
  Magazine}, vol.~52, no.~12, pp. 110--121, 2014.

\bibitem{haneda2015channel}
K.~Haneda, ``Channel models and beamforming at millimeter-wave frequency
  bands,'' \emph{IEICE Transactions on Communications}, vol.~98, no.~5, pp.
  755--772, May 2015.

\bibitem{Rap15a}
T.~S. Rappaport \emph{et~al.}, \emph{Millimeter Wave Wireless
  Communications}.\hskip 1em plus 0.5em minus 0.4em\relax Pearson/Prentice
  Hall, 2015.

\bibitem{yamada2015experimental}
T.~Yamada \emph{et~al.}, ``Experimental evaluation of ieee 802.11ad
  millimeter-wave wlan devices,'' in \emph{2015 21st Asia-Pacific Conference on
  Communications (APCC)}, Oct. 2015, pp. 278--282.

\bibitem{siligaris2011a}
A.~Siligaris \emph{et~al.}, ``A 65-nm {CMOS} fully integrated transceiver
  module for 60-{GHz} {Wireless HD} applications,'' \emph{IEEE Journal of
  Solid-State Circuits}, vol.~46, no.~12, pp. 3005--3017, Dec. 2011.

\bibitem{charfi2013phy}
E.~Charfi, L.~Chaari, and L.~Kamoun, ``{PHY}/{MAC} enhancements and qos
  mechanisms for very high throughput {WLANs}: A survey,'' \emph{IEEE
  Communications Surveys Tutorials}, vol.~15, no.~4, pp. 1714--1735, Apr. 2013.

\bibitem{perahia2013next}
E.~Perahia and R.~Stacey, \emph{Next Generation Wireless LANS: 802.11 n and
  802.11 ac}.\hskip 1em plus 0.5em minus 0.4em\relax Cambridge university
  press, 2013.

\bibitem{verma2013wifi}
L.~Verma \emph{et~al.}, ``Wifi on steroids: 802.11 ac and 802.11 ad,''
  \emph{IEEE Wireless Communications}, vol.~20, no.~6, pp. 30--35, Dec. 2013.

\bibitem{perahia2011gigabit}
E.~Perahia and M.~X. Gong, ``Gigabit wireless {LANs}: an overview of {IEEE}
  802.11 ac and 802.11 ad,'' \emph{ACM SIGMOBILE Mobile Computing and
  Communications Review}, vol.~15, no.~3, pp. 23--33, Nov. 2011.

\bibitem{perahia2010ieee}
E.~Perahia \emph{et~al.}, ``{IEEE} 802.11ad: Defining the next generation
  multi-{Gbps} {Wi-Fi},'' in \emph{2010 7th IEEE Consumer Communications and
  Networking Conference}, Jan. 2010, pp. 1--5.

\bibitem{802.11ad}
\BIBentryALTinterwordspacing
A.~Maltsev \emph{et~al.}, ``\textcolor{black}{{Channel Models for 60 GHz WLAN
  Systems}},'' Tech. Rep. doc.: IEEE 802.11-09/0334r8, May 2010. [Online].
  Available:
  \url{https://mentor.ieee.org/802.11/dcn/13/11-13-0794-00-00aj-channel-models-for-45-ghz-wlan-systems.docx}
\BIBentrySTDinterwordspacing

\bibitem{maltsev2016channel}
------, ``{Channel modeling in the next generation mmWave Wi-Fi: IEEE 802.11ay
  standard},'' in \emph{European Wireless 2016; 22th European Wireless
  Conference}, May 2016, pp. 1--8.

\bibitem{802.11p}
{IEEE 802.11 Working Group} \emph{et~al.}, ``\textcolor{black}{IEEE Standard
  for Information technology-Telecommunications and information exchange
  between systems-Local and metropolitan area networks-Specific requirements
  Part 11: Wireless LAN Medium Access Control (MAC) and Physical Layer (PHY)
  Specifications},'' \emph{IEEE Std}, vol. 802, no.~11, Sept. 2010.

\bibitem{moustafa2009vehicular}
H.~Moustafa and Y.~Zhang, \emph{\textcolor{black}{Vehicular networks:
  techniques, standards, and applications}}.\hskip 1em plus 0.5em minus
  0.4em\relax Auerbach publications, 2009.

\bibitem{bendor2011mmwave}
E.~Ben-Dor \emph{et~al.}, ``\textcolor{black}{Millimeter-Wave 60 {GHz} Outdoor
  and Vehicle {AOA} Propagation Measurements Using a Broadband Channel
  Sounder},'' in \emph{2011 IEEE Global Telecommunications Conference -
  GLOBECOM 2011}, Dec. 2011, pp. 1--6.

\bibitem{rappaport2010analysis}
T.~S. Rappaport, S.~DiPierro, and R.~Akturan, ``\textcolor{black}{Analysis and
  simulation of adjacent service interference to vehicle-equipped digital
  wireless receivers from cellular mobile terminals}s,'' in \emph{Vehicular
  Technology Conference Fall (VTC 2010-Fall), 2010 IEEE 72nd}, Sept. 2010, pp.
  1--5.

\bibitem{rap2016ap}
T.~S. Rappaport \emph{et~al.}, ``{Small-Scale, Local Area, and Transitional
  Millimeter Wave Propagation for 5G Cellular Communications},'' \emph{IEEE
  Transactions on Antennas and Propagation}, this issue.

\bibitem{shivaldova2012roadside}
V.~Shivaldova \emph{et~al.}, ``\textcolor{black}{On roadside unit antenna
  measurements for vehicle-to-infrastructure communications},'' in
  \emph{Personal Indoor and Mobile Radio Communications (PIMRC), 2012 IEEE 23rd
  International Symposium on}, Sept. 2012, pp. 1295--1299.

\bibitem{phan2015making}
D.-T. Phan-Huy, M.~Sternad, and T.~Svensson, ``\textcolor{black}{Making 5G
  adaptive antennas work for very fast moving vehicles},'' \emph{IEEE
  Intelligent Transportation Systems Magazine}, vol.~7, no.~2, pp. 71--84,
  Summer 2015.

\bibitem{rappaport2011state}
T.~S. Rappaport, J.~N. Murdock, and F.~Gutierrez, ``State of the art in
  60-{GHz} integrated circuits and systems for wireless communications,''
  \emph{Proceedings of the IEEE}, vol.~99, no.~8, pp. 1390--1436, Aug. 2011.

\bibitem{ITU-Rattenuation}
ITU-R, ``\textcolor{black}{Attenuation by Atmospheric Gases},'' Tech. Rep.
  P.676-11, Sept. 2016.

\bibitem{sun2017a}
S.~Sun, G.~R. {MacCartney, Jr.}, and T.~S. Rappaport, ``\textcolor{black}{A
  Novel Millimeter-Wave channel simulator and applications for 5G wireless
  communications},'' in \emph{IEEE International Conference on Communication
  (ICC)}, May 2017, pp. 1--7.

\bibitem{xu2000measurements}
H.~Xu \emph{et~al.}, ``Measurements and models for 38-{GHz} point-to-multipoint
  radiowave propagation,'' \emph{IEEE Journal on Selected Areas in
  Communications}, vol.~18, no.~3, pp. 310--321, Mar. 2000.

\bibitem{ITU-Rspecific}
ITU-R, ``Specific attenuation model for rain for use in prediction methods,
  propagation in non-ionized media,'' Tech. Rep. P.838-3, 2005.

\bibitem{sun2016propagation}
S.~Sun \emph{et~al.}, ``{Propagation Path Loss Models for 5G Urban Micro- and
  Macro-Cellular Scenarios},'' in \emph{2016 IEEE 83rd Vehicular Technology
  Conference (VTC 2016-Spring)}, May 2016, pp. 1--6.

\bibitem{sun2015path}
------, ``Path loss, shadow fading, and line-of-sight probability models for
  {5G} urban macro-cellular scenarios,'' in \emph{2015 IEEE Globecom Workshops
  (GC Wkshps)}, Dec. 2015, pp. 1--7.

\bibitem{friis1946note}
H.~T. Friis, ``A note on a simple transmission formula,'' \emph{Proceedings of
  the IRE}, vol.~34, no.~5, pp. 254--256, May 1946.

\bibitem{uchendu2016survey}
I.~Uchendu and J.~R. Kelly, ``\textcolor{black}{Survey of beam steering
  techniques available for millimeter wave applications},'' \emph{Progress In
  Electromagnetics Research B}, vol.~68, pp. 35--54, Mar. 2016.

\bibitem{nitsche2015steering}
T.~Nitsche \emph{et~al.}, ``\textcolor{black}{Steering with eyes closed:
  mm-wave beam steering without in-band measurement},'' in \emph{Computer
  Communications (INFOCOM), 2015 IEEE Conference on}, Apr. 2015, pp.
  2416--2424.

\bibitem{maccartney2016millimeter}
G.~R. {MacCartney, Jr.} \emph{et~al.}, ``{Millimeter-Wave Human Blockage at 73
  GHz with a Simple Double Knife-Edge Diffraction Model and Extension for
  Directional Antennas},'' in \emph{IEEE 84th Vehicular Technology Conference
  Fall (VTC 2016-Fall)}, Sept. 2016, pp. 1--6.

\bibitem{rodriguez2015analysis}
I.~Rodriguez \emph{et~al.}, ``Analysis of 38 {GHz} mmwave propagation
  characteristics of urban scenarios,'' in \emph{European Wireless 2015; 21th
  European Wireless Conference; Proceedings of}, May 2015, pp. 1--8.

\bibitem{jacque2016indoor}
J.~Ryan, G.~R. {MacCartney, Jr.}, and T.~S. Rappaport, ``{Indoor Office
  Wideband Penetration Loss Measurements at 73 GHz},'' in \emph{IEEE
  International Conference on Communications Workship (ICCW)}, May 2017.

\bibitem{rumney2016testing2}
M.~Rumney, ``{Testing 5G: Time to throw away the cables},'' \emph{Microwave
  Journal}, Nov. 2016.

\bibitem{mmMAGIC}
\BIBentryALTinterwordspacing
mmMAGIC, ``Measurement results and final mmmagic channel models,'' Tech. Rep.
  H2020-ICT-671650-mmMAGIC/D2.2, May 2017. [Online]. Available:
  \url{https://5g-mmmagic.eu/results/}
\BIBentrySTDinterwordspacing

\bibitem{samimi201628}
M.~K. Samimi \emph{et~al.}, ``28 {GHz} millimeter-wave ultrawideband
  small-scale fading models in wireless channels,'' in \emph{2016 IEEE 83rd
  Vehicular Technology Conference (VTC 2016-Spring)}, May 2016, pp. 1--6.

\bibitem{samimi2016local}
M.~K. Samimi and T.~S. Rappaport, ``{Local multipath model parameters for
  generating 5G millimeter-wave 3GPP-like channel impulse response},'' in
  \emph{2016 10th European Conference on Antennas and Propagation (EuCAP)},
  Apr. 2016, pp. 1--5.

\bibitem{sijia2016}
S.~Deng \emph{et~al.}, ``{Indoor and Outdoor 5G Diffraction Measurements and
  Models at 10, 20, and 26 GHz},'' in \emph{2016 IEEE Global Communications
  Conference (GLOBECOM)}, Dec. 2016, pp. 1--7.

\bibitem{3GPP2014}
\BIBentryALTinterwordspacing
3GPP, ``Technical specification group radio access network; study on 3d channel
  model for lte (release 12),'' 3rd Generation Partnership Project (3GPP), TR
  36.873 V12.2.0, June 2015. [Online]. Available:
  \url{http://www.3gpp.org/dynareport/36873.htm}
\BIBentrySTDinterwordspacing

\bibitem{ertel1998overview}
R.~B. Ertel \emph{et~al.}, ``Overview of spatial channel models for antenna
  array communication systems,'' \emph{IEEE personal communications}, vol.~5,
  no.~1, pp. 10--22, Feb. 1998.

\bibitem{sun2017b}
S.~Sun \emph{et~al.}, ``Millimeter wave small-scale spatial statistics in an
  urban microcell scenario,'' in \emph{IEEE International Conference on
  Communication (ICC)}, May 2017, pp. 1--7.

\bibitem{rappaport20175g}
T.~S. Rappaport, S.~Sun, and M.~Shafi, ``{5G} channel model with improved
  accuracy and efficiency in mmwave bands,'' \emph{IEEE 5G Tech Focus}, Mar.
  2017.

\bibitem{rappaport2017VTC}
------, ``Investigation and comparison of {3GPP} and {NYUSIM} channel models
  for {5G} wireless communications,'' \emph{2017 IEEE 86th Vehicular Technology
  Conference (VTC Fall)}, Sept. 2017.

\bibitem{3GPP2017}
\BIBentryALTinterwordspacing
3GPP, ``Study on channel model for frequencies from 0.5 to 100 {GHz},'' 3rd
  Generation Partnership Project (3GPP), TR 38.901 V14.0.0, May. 2017.
  [Online]. Available: \url{http://www.3gpp.org/DynaReport/38901.htm}
\BIBentrySTDinterwordspacing

\bibitem{METIS2015}
\BIBentryALTinterwordspacing
METIS2020, ``{METIS Channel Model},'' Tech. Rep. METIS2020, Deliverable D1.4
  v3, July 2015. [Online]. Available:
  \url{https://www.metis2020.com/wp-content/uploads/deliverables/METIS\_D1.4\_v3.0.pdf}
\BIBentrySTDinterwordspacing

\bibitem{samimi20153}
M.~K. Samimi and T.~S. Rappaport, ``{3-D} statistical channel model for
  millimeter-wave outdoor mobile broadband communications,'' in \emph{IEEE
  International Conference on Communications (ICC)}, June 2015, pp. 2430--2436.

\bibitem{MiWEBA}
\BIBentryALTinterwordspacing
MiWeba, ``{WP5: Propagation, Antennas and Multi-Antenna Technique; D5.1:
  Channel Modeling and Characterization},'' Tech. Rep. {MiWEBA Deliverable
  D5.1}, June 2014. [Online]. Available:
  \url{http://www.miweba.eu/wp-content/uploads/2014/07/MiWEBA D5.1 v1.011.pdf}
\BIBentrySTDinterwordspacing

\bibitem{hur2016proposal}
S.~Hur \emph{et~al.}, ``{Proposal on Millimeter-Wave Channel Modeling for 5G
  Cellular System},'' \emph{IEEE Journal of Selected Topics in Signal
  Processing}, vol.~10, no.~3, pp. 454--469, Apr. 2016.

\bibitem{ITU-RM.2135}
{International Telecommunications Union}, ``Guidelines for evaluation of radio
  interface technologies for {IMT-Advanced},'' Geneva, Switzerland, REP. ITU-R
  M.2135-1, Dec. 2009.

\bibitem{jarvelainen2016evaluation}
J.~Jarvelainen \emph{et~al.}, ``Evaluation of millimeter-wave line-of-sight
  probability with point cloud data,'' \emph{IEEE Wireless Communications
  Letters}, vol.~5, no.~3, pp. 228--231, June 2016.

\bibitem{piersanti2012millimeter}
S.~Piersanti, L.~A. Annoni, and D.~Cassioli, ``Millimeter waves channel
  measurements and path loss models,'' in \emph{2012 IEEE International
  Conference on Communications (ICC)}, June 2012, pp. 4552--4556.

\bibitem{semaan2014outdoor}
E.~Semaan \emph{et~al.}, ``Outdoor-to-indoor coverage in high frequency
  bands,'' in \emph{2014 IEEE Globecom Workshops (GC Wkshps)}, Dec. 2014, pp.
  393--398.

\bibitem{maccartney2015millimeter}
G.~R. {MacCartney, Jr.} \emph{et~al.}, ``\textcolor{black}{Millimeter-wave
  omnidirectional path loss data for small cell 5G channel modeling},''
  \emph{IEEE Access}, vol.~3, pp. 1573--1580, Aug. 2015.

\bibitem{maccartney2017study}
G.~R. {MacCartney, Jr.} and T.~S. Rappaport, ``\textcolor{black}{Study on 3GPP
  rural macrocell path loss models for millimeter wave wireless
  communications},'' in \emph{2017 IEEE International Conference on
  Communications (ICC)}, May 2017, pp. 1--7.

\bibitem{3GPP2016}
\BIBentryALTinterwordspacing
3GPP, ``Technical specification group radio access network; channel model for
  frequency spectrum above 6 {GHz},'' 3rd Generation Partnership Project
  (3GPP), TR 38.900 V14.2.0, Dec. 2016. [Online]. Available:
  \url{http://www.3gpp.org/DynaReport/38900.htm}
\BIBentrySTDinterwordspacing

\bibitem{WINNERplus}
{WINNER+ D5.3}, ``Final channel models,'' Tech. Rep. V1.0, CELTIC CP5-026
  WINNER+ project, 2010.

\bibitem{hata1980}
M.~Hata, ``Empirical formula for propagation loss in land mobile radio
  services,'' \emph{IEEE Transactions on Vehicular Technology}, vol.~29, no.~3,
  pp. 317--325, Aug. 1980.

\bibitem{sun2015synthesizing}
S.~Sun \emph{et~al.}, ``{Synthesizing Omnidirectional antenna patterns,
  received power and path loss from directional antennas for 5G millimeter-wave
  communications},'' in \emph{IEEE Global Communications Conference
  (GLOBECOM)}, Dec. 2015, pp. 3948--3953.

\bibitem{maccartney2014primrc}
G.~R. MacCartney, M.~K. Samimi, and T.~S. Rappaport, ``Omnidirectional path
  loss models in new york city at 28 ghz and 73 ghz,'' in \emph{2014 IEEE 25th
  Annual International Symposium on Personal, Indoor, and Mobile Radio
  Communication (PIMRC)}, Sept 2014, pp. 227--231.

\bibitem{andersen1995}
J.~B. Andersen, T.~S. Rappaport, and S.~Yoshida, ``Propagation measurements and
  models for wireless communications channels,'' \emph{IEEE Communications
  Magazine}, vol.~33, no.~1, pp. 42--49, Jan. 1995.

\bibitem{bullington1947radio}
K.~Bullington, ``Radio propagation at frequencies above 30 megacycles,''
  \emph{Proceedings of the IRE}, vol.~35, no.~10, pp. 1122--1136, Oct. 1947.

\bibitem{feuerstein1994path}
M.~J. Feuerstein \emph{et~al.}, ``Path loss, delay spread, and outage models as
  functions of antenna height for microcellular system design,'' \emph{IEEE
  Transactions on Vehicular Technology}, vol.~43, no.~3, pp. 487--498, Aug.
  1994.

\bibitem{3GPPTDOC}
3GPP, ``Correction for low and high frequency model harmonization,'' Ericsson,
  Samsung, NTT DOCOMO, Nokia, Intel, Telstra, Tech. Rep. TDOC R1-1701195, Jan.
  2017.

\bibitem{TDOC164975}
{3GPP}, ``New measurements at 24 {GHz} in a rural macro environment,'' Telstra,
  Ericsson, Tech. Rep. TDOC R1-164975, May 2016.

\bibitem{rajendran2011concepts}
V.~K. Rajendran \emph{et~al.}, ``\textcolor{black}{Concepts and Implementation
  of a Semantic Web Archiving and Simulation System for RF Propagation
  Measurements},'' in \emph{Vehicular Technology Conference (VTC 2011-Fall),
  2011 IEEE}, Sept. 2011, pp. 1--5.

\bibitem{jaeckel2014quadriga}
S.~Jaeckel \emph{et~al.}, ``\textcolor{black}{Quadriga: A 3-d multi-cell
  channel model with time evolution for enabling virtual field trials},''
  \emph{IEEE Transactions on Antennas and Propagation}, vol.~62, no.~6, pp.
  3242--3256, Mar. 2014.

\bibitem{NYUSIM}
\BIBentryALTinterwordspacing
{New York University}, \emph{\textcolor{black}{NYUSIM}}, 2016. [Online].
  Available: \url{http://wireless.engineering.nyu.edu/nyusim/}
\BIBentrySTDinterwordspacing

\bibitem{HUAWEI}
\BIBentryALTinterwordspacing
{Huawei}, ``{5G: Huawei and Vodafone achieve 20Gbps for single-user outdoor at
  E-Band},'' July 2016. [Online]. Available:
  \url{http://www.huawei.com/en/news/2016/7/huawei-vodafone-5g-test}
\BIBentrySTDinterwordspacing

\bibitem{ERICSSON}
\BIBentryALTinterwordspacing
{Ericsson}, ``{Ericsson and Telstra conduct the first live 5G trial in
  Australia},'' Sept. 2016. [Online]. Available:
  \url{https://www.ericsson.com/news/160920-ericsson-telstra-5g-trial-australia\_244039854\_c}
\BIBentrySTDinterwordspacing

\end{thebibliography}
\end{document}